\documentclass[aps,prl,reprint,superscriptaddress,nofootinbib,longbibliography,floatfix]{revtex4-2}
\usepackage{amsmath,amssymb,bm,graphicx,hyperref}
\hypersetup{colorlinks=true,linkcolor=blue,urlcolor=blue,citecolor=blue}

\newcommand{\runinhead}[1]{\par\indent\textit{#1}\textemdash\allowbreak\ignorespaces}

\begin{document}
\title{Thermally Reentrant Charge-\texorpdfstring{$2e$}{2e} Coherence in a Charge-\texorpdfstring{$4e$}{4e} Superconductor}

\author{Yan-Guang Yue}
\affiliation{Department of Physics and State Key Laboratory of Surface Physics, Fudan University, Shanghai 200433, China}
\author{Jie Lou}
\affiliation{Department of Physics and State Key Laboratory of Surface Physics, Fudan University, Shanghai 200433, China}
\author{Yan Chen}
\email{yanchen99@fudan.edu.cn}
\affiliation{Department of Physics and State Key Laboratory of Surface Physics, Fudan University, Shanghai 200433, China}
\affiliation{Shanghai Branch, Hefei National Laboratory, Shanghai 201315, China}

\begin{abstract}

Heating can restore Cooper-pair coherence after it has been lost. We find this reversal in a two-dimensional generalized XY model with annealed two-state bond variables. For a range of microscopic parameters, a heating path at fixed couplings passes through two charge-$4e$ regimes separated by a reentrant charge-$2e$ superconducting interval. In the constrained limit, an exact Ising mapping relates recovery to the entropy gained through selective thermal activation of aligned bonds and gives a necessary condition for reentrance. Monte Carlo simulations at finite coupling find restored algebraic charge-$2e$ correlations and scaling compatible with the Ising thermal exponent. The free energy of a system-spanning domain wall is consistent with renewed half-vortex confinement in the reentrant superconducting regime.

\end{abstract}
\maketitle

\runinhead{Introduction.}
Thermal melting can proceed in stages~\cite{HalperinNelson1978,NelsonHalperin1979,Young1979}. In correlated systems, fluctuations can suppress primary order while preserving order in a composite field, giving rise to vestigial phases~\cite{Chandra1990,Kivelson1998,Fernandes2012,FernandesOrthSchmalian2019}. Charge-$4e$ superconductivity provides an example. A composite of two charge-$2e$ superconducting fields can retain phase coherence after conventional Cooper-pair coherence is lost. Its stability has been studied in pair-density-wave superconductors~\cite{Agterberg2008,Berg2009,Agterberg2020,WuWang2024,HueckerWang2026,HueckerWangErratum2026} and their neutral superfluid analogues~\cite{RadzihovskyVishwanath2009,Radzihovsky2011}, in multicomponent superconductors~\cite{FernandesFu2021,Jian2021,Hecker2023,HeckerFernandes2024,Liu2024,Zou2026,Maccari2026}, and in Josephson arrays and coupled superconducting layers~\cite{DoucotVidal2002,SongZhang2022,Liu2023}. Experiments have reported $h/4e$ and $h/6e$ flux periodicities in kagome superconducting rings~\cite{Ge2024} and coherent charge-$4e$ transport in Josephson devices~\cite{Gladchenko2009,Huang2022,Ciaccia2024,Leblanc2024}.

In two dimensions, vortex unbinding destroys algebraic phase coherence in the XY model at a Berezinskii--Kosterlitz--Thouless (BKT) transition~\cite{Berezinskii1971,KT1973,Kosterlitz1974}. Generalized XY models with higher-harmonic interactions support paired phases, half vortices, and domain walls~\cite{LeeGrinstein1985,Korshunov1986,CarpenterChalker1989,Poderoso2011,Shi2011,HubscherWessel2013,Serna2017}. A half vortex carries a $\pi$ winding of the smooth charge-$2e$ phase and a $2\pi$ winding of the charge-$4e$ phase. A domain wall with positive tension confines half vortices linearly. The loss of this tension can destroy charge-$2e$ coherence while charge-$4e$ correlations remain algebraic. Half vortices can then remain logarithmically bound until their unbinding at a higher-temperature BKT transition destroys charge-$4e$ coherence~\cite{Shi2011,Serna2017,SongZhang2022,Zou2026}.

Heating can also restore order. Reentrant superconductivity is known in systems with competing magnetic and superconducting orders~\cite{Fertig1977,WuValls2012}. Internal degrees of freedom can favor order through entropy, as in inverse melting~\cite{Schupper2004}. In decorated Ising models, tracing over internal spins~\cite{Fisher1959} can give rise to reentrant magnetic order~\cite{SyoziNakano1968,Matasovska2007}. Once charge-$2e$ coherence has melted, can further heating restore it and reestablish linear confinement of half vortices while charge-$4e$ coherence persists?

In this Letter, we study a generalized XY model on a triangular lattice. Each bond carries an annealed two-state variable. For a range of microscopic parameters, heating at fixed couplings produces the sequence
\begin{equation}
 \mathrm{SC}\longrightarrow\text{charge-}4e\longrightarrow\mathrm{SC}
 \longrightarrow\text{charge-}4e\longrightarrow\mathrm{N},
 \label{eq:sequence}
\end{equation}
where SC denotes conventional charge-$2e$ superconductivity and N the normal phase. In the limit of infinite second-harmonic coupling $J_2$, the model maps exactly onto a ferromagnetic Ising model. Its two spin states represent charge-$2e$ phase choices differing by $\pi$. Selective thermal activation of aligned bonds restores Ising branch order. The triangular-lattice Ising critical coupling $K_c$ satisfies the necessary entropy bound $2K_c<\ln 2$. On the square lattice $2K_c$ exceeds $\ln 2$. At finite $J_2$, Monte Carlo simulations find that charge-$2e$ correlations recover algebraic decay between the two charge-$4e$ intervals. Charge-$4e$ correlations remain algebraic throughout all four superconducting regimes. Scaling at the recovery boundary is compatible with the Ising thermal exponent. We also measure the free-energy cost of a system-spanning domain wall in the vortex-free, zero-winding sector of the charge-$4e$ field. Its growth with system size resumes in the reentrant SC interval.

\begin{figure}[!tb]
 \centering
 \includegraphics[width=\columnwidth]{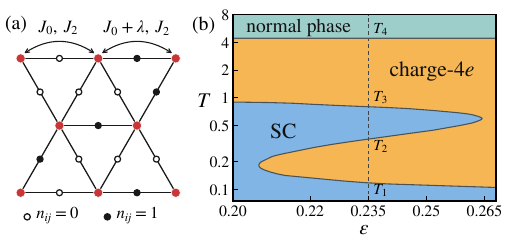}
 \caption[Model and reentrant phase diagram]{Model and phase diagram. (a) Sites carry $U(1)$ phase fields $\theta_i$. Open and filled bond centers denote $n_{ij}=0$ and $1$, respectively. The coupling favoring phase alignment is $J_0$ on an unexcited bond and $J_0+\lambda$ on an excited bond. The $J_2$ coupling acts in both states. Equation~\eqref{eq:model} gives the bond excitation energy. (b) Phase diagram for $J_0=0.025$, $\lambda=0.975$, and $J_2=3$. Both charge-$2e$ and charge-$4e$ correlations are algebraic in SC. Only charge-$4e$ correlations remain algebraic in the charge-$4e$ phase. Both are short ranged in the normal phase. Heating along the dashed line at $\varepsilon=0.235$ gives the sequence SC--charge-$4e$--SC--charge-$4e$--N. The curves summarize finite-size boundary estimates. }
 \label{fig:phase}
\end{figure}

\runinhead{Model and phase diagram.}
We consider a generalized XY model on the triangular lattice shown in Fig.~\ref{fig:phase}(a). The sites carry $U(1)$ phase fields $\theta_i\in[0,2\pi)$, and $n_{ij}\in\{0,1\}$ labels the state of each nearest-neighbor bond. The Hamiltonian is
\begin{align}
 H=\sum_{\langle ij\rangle}\Big[&-J_0\cos\theta_{ij}-J_2\cos2\theta_{ij}\nonumber\\
 &+n_{ij}\{\varepsilon+\lambda(1-\cos\theta_{ij})\}\Big],
 \label{eq:model}
\end{align}
where $\theta_{ij}=\theta_i-\theta_j$ and $J_0,J_2,\varepsilon,\lambda>0$.

The $J_2$ interaction favors relative phases of either $0$ or $\pi$, whereas the coupling $J_0+\lambda n_{ij}$ favors phase alignment. Exciting a bond from $n_{ij}=0$ to $1$ strengthens the latter coupling from $J_0$ to $J_0+\lambda$. The bond excitation energy is $\varepsilon+\lambda(1-\cos\theta_{ij})$.

For these positive couplings, the minimum bond energy $-J_0-J_2$ occurs at $n_{ij}=0$ and $\theta_{ij}=0$. All bonds can attain this minimum simultaneously, giving a uniform ground-state phase $\theta_i=\theta_0$. We identify $\theta_i$ with the phase of a fixed-amplitude charge-$2e$ field, so this ground state represents SC.

At finite temperature, superconducting coherence in two dimensions is characterized by algebraic phase correlations~\cite{Mermin1966}. Correlations of $e^{i\theta_i}$ probe charge-$2e$ coherence, while those of the charge-$4e$ composite field $e^{2i\theta_i}$ probe charge-$4e$ coherence. Both correlations are algebraic in SC. In the charge-$4e$ phase, charge-$2e$ correlations are short ranged while charge-$4e$ correlations remain algebraic. Both are short ranged in the normal phase.

Monte Carlo simulations give the finite-size estimates of the phase boundaries in Fig.~\ref{fig:phase}(b) for $J_0=0.025$, $\lambda=0.975$, and $J_2=3$. We use $J_0+\lambda=1$ as the energy unit and set $k_B=1$. We vary $\varepsilon$ between temperature scans. Along each heating path, all microscopic parameters, including $\varepsilon$, remain fixed.

Along $\varepsilon=0.235$, charge-$2e$ coherence first melts near $T_1\simeq0.118$, leaving the charge-$4e$ phase. Further heating restores charge-$2e$ coherence near $T_2\simeq0.354$, and the system reenters SC. The reentrant SC interval ends near $T_3\simeq0.798$, where the charge-$4e$ phase reappears. Charge-$4e$ coherence finally melts near $T_4\simeq4.44$, above which the system is normal.

As $\varepsilon$ decreases, the lower charge-$4e$ window narrows. Increasing $\varepsilon$ narrows the reentrant SC window. The estimated closures in Fig.~\ref{fig:phase}(b) delimit the range in which both windows are resolved. Finite-size scans at $J_2=2$ and $4$ also show crossing features at each of the three SC--charge-$4e$ boundaries [Supplemental Material (SM)~\cite{SupplementalMaterial}].

\begingroup\clubpenalty=10000\relax
\runinhead{Thermally reentrant Ising order in the constrained limit.}
For exactly aligned neighbors ($\theta_{ij}=0$), exciting the bond costs $\varepsilon$. For antialigned neighbors ($\theta_{ij}=\pi$), it costs $\varepsilon+2\lambda$. The corresponding conditional excitation probabilities are
\begin{equation}
 p_+(T)=\frac{1}{1+e^{\varepsilon/T}},\qquad
 p_-(T)=\frac{1}{1+e^{(\varepsilon+2\lambda)/T}}.
 \label{eq:activation}
\end{equation}
\par\endgroup
At intermediate temperatures, aligned bonds are much more likely to be excited than antialigned bonds. Summing over the bond variable therefore gives aligned configurations a larger entropic weight.

\begin{figure}[!htbp]
 \centering
 \includegraphics[width=\columnwidth]{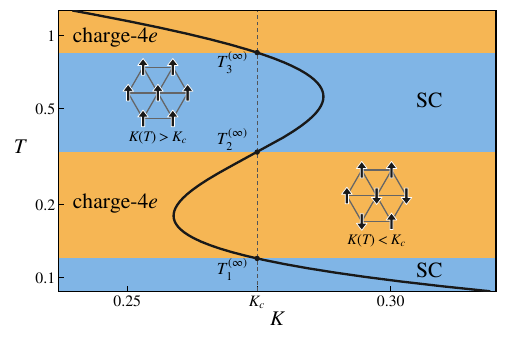}
 \caption[Thermal reentrance in the constrained model]{Thermal reentrance in the strictly constrained triangular-lattice model.
The dark curve gives $K(T)=J_{\mathrm I}(T)/T$ for $J_2\to\infty$,
$J_0=0.025$, $\lambda=0.975$, and $\varepsilon=0.235$, with $k_B=1$.
On heating, the curve crosses the Ising threshold $K_c=\ln(3)/4$
(vertical dashed line) at $T_1^{(\infty)}$, $T_2^{(\infty)}$, and
$T_3^{(\infty)}$. Each horizontal band identifies the phase along the curve
in that temperature interval. In the right inset, mixed arrows illustrate the loss of Ising branch order in the
intermediate charge-$4e$ phase ($K<K_c$). In the left inset, aligned arrows
illustrate an idealized configuration with Ising branch order in the reentrant SC interval ($K>K_c$).
Up and down arrows denote the two charge-$2e$ phase branches $\theta_0$ and
$\theta_0+\pi$, respectively. The charge-$4e$ composite field remains spatially uniform throughout.
The horizontal scale shows the vicinity of $K_c$. }
 \label{fig:mechanism}
\end{figure}

For $J_2\to\infty$ at fixed $T$, the phase of the charge-$4e$ composite field locks across the lattice, so $e^{2i\theta_i}=e^{2i\theta_0}$ at every site. The normalized charge-$2e$ field takes the form $e^{i\theta_i}=\sigma_i e^{i\theta_0}$, with $\sigma_i=\pm1$. The two signs represent local phase choices differing by $\pi$ (Fig.~\ref{fig:mechanism}). Summing over the bond states gives an effective Hamiltonian for the Ising spins,
\begin{equation}
 \begin{aligned}
 H_{\mathrm I}(T)&=-J_{\mathrm I}(T)\sum_{\langle ij\rangle}\sigma_i\sigma_j,\\
 K(T)\equiv\frac{J_{\mathrm I}(T)}{T}
 &=\frac{J_0}{T}+\frac12\ln\frac{1+e^{-\varepsilon/T}}
 {1+e^{-(\varepsilon+2\lambda)/T}} .
 \end{aligned}
 \label{eq:effectiveK}
\end{equation}
The effective Ising energy coupling is ferromagnetic ($J_{\mathrm I}(T)>0$) at every temperature. Its ratio to the thermal energy, $K(T)$, is the dimensionless coupling that controls Ising branch order.

For any two sites, the charge-$2e$ and charge-$4e$ correlations satisfy
\begin{equation}
 \langle e^{i\theta_{ij}}\rangle=\langle\sigma_i\sigma_j\rangle_{\mathrm I},
 \qquad \langle e^{2i\theta_{ij}}\rangle=1 ,
 \label{eq:constrainedCorrelations}
\end{equation}
where the subscript $\mathrm I$ denotes an average in the effective Ising model. The charge-$2e$ correlation thus measures Ising branch order against a uniform charge-$4e$ background. On the triangular lattice the exact ordering threshold is $K_c=\tfrac14\ln 3$~\cite{Houtappel1950}. For $J_0=0.025$, $\lambda=0.975$, and $\varepsilon=0.235$, the three crossings in Fig.~\ref{fig:mechanism} occur at
\begin{equation}
 \bigl(T_1^{(\infty)},T_2^{(\infty)},T_3^{(\infty)}\bigr)
 \simeq (0.12,\,0.33,\,0.85).
 \label{eq:constrainedRoots}
\end{equation}
The constrained model therefore has two SC intervals separated by a charge-$4e$ interval and followed by a second charge-$4e$ interval.

As $T\to0$, bond excitations are suppressed and $K(T)\sim J_0/T$. Heating initially lowers $K(T)$ through $K_c$. To restore Ising branch order, the free-energy advantage $2J_{\mathrm I}(T)$ of an aligned bond must rise through the threshold $2TK_c$. Its slope is the entropy difference between aligned and antialigned bonds (SM). At $T_2^{(\infty)}$, $\mathrm{d}(2J_{\mathrm I})/\mathrm{d}T\simeq0.62>2K_c\simeq0.55$. A two-state bond supplies at most $\ln 2$ of entropy, so an upward crossing requires $2K_c<\ln 2$. Whether it occurs depends on the microscopic parameters through $K(T)$.

At higher temperatures, excitations on antialigned bonds become common. The entropy difference shrinks, and $K(T)$ falls below $K_c$ for the last time. As $T\to\infty$, $J_{\mathrm I}(T)$ approaches $J_0+\lambda/2$ and $K(T)$ vanishes. All three crossings have nonzero slope, so they persist under small changes of the microscopic parameters. The SM gives the full constrained critical curve.

In the finite-$J_2$ model the charge-$4e$ phase fluctuates and can itself melt. We use correlations and finite-size scaling to identify the phases in this regime.

\runinhead{Recovery of charge-$2e$ coherence at finite coupling.}
We test for recovery of charge-$2e$ coherence in the finite-$J_2$ model by Monte Carlo sampling on periodic triangular lattices at $J_2=3$ and $\varepsilon=0.235$. We measure the spatial correlation function and the zero-momentum structure factor in each charge channel,
\begin{align}
 G_{2me}(r)&=\frac{1}{3N_s}\sum_{i,\mu}
 \big\langle\cos[m(\theta_i-\theta_{i+r\mathbf{d}_\mu})]\big\rangle,
 \label{eq:correlations}\\
 S_{2me}(L,T)&=\frac{1}{N_s}\left\langle\left|\sum_{j=1}^{N_s}e^{im\theta_j}\right|^2\right\rangle.
 \label{eq:structure}
\end{align}
Here $m=1,2$ labels the charge-$2e$ and charge-$4e$ channels, respectively, $N_s=L^2$, and $\mathbf{d}_\mu$ runs over three equivalent nearest-neighbor directions. At fixed $T$, algebraic correlations $G_{2me}(r)\sim r^{-\eta_{2me}}$ give $S_{2me}(L)\sim L^{2-\eta_{2me}}$. For short-range correlations, $S_{2me}$ saturates once $L$ exceeds the correlation length.

\begin{figure}[!t]
 \centering
 \includegraphics[width=\columnwidth]{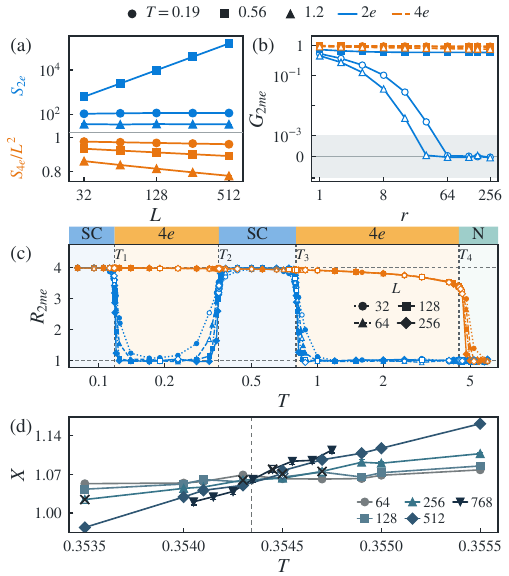}
 \caption[Charge correlations and the recovery transition]{Charge correlations at $J_2=3$ and $\varepsilon=0.235$. Blue and orange denote the $2e$ and $4e$ channels. (a) Size dependence of $S_{2e}$ (top) and $S_{4e}/L^2$ (bottom). Points pass the equilibration and sampling checks for the plotted observable. (b) Signed spatial correlations at $L=512$. Shapes in (a,b) mark temperature. Open symbols in (b) mark histories with an unresolved $2e$ correlation ratio. The vertical axis is linear in the shaded interval $|G_{2me}|\le10^{-3}$. (c) Size-doubling ratios. The color bar above the plot shows the heating sequence SC--charge-$4e$--SC--charge-$4e$--N, in the colors of Fig.~\ref{fig:phase}(b). Dashed vertical lines mark $T_1$--$T_4$. Shapes and line styles identify the smaller size $L$. Open symbols mark estimates with an equilibration warning, a low effective sample size, or only one usable cutoff. (d) Recovery crossing at the later cutoff. Symbols identify $L$. Crosses mark five excluded points. The dashed line gives $T_2$ from the 40 accepted points at $64\le L\le768$. Error bars show one standard error. In (c) we propagate the independent structure-factor errors at $L$ and $2L$. See SM, Secs.~S4 and S5, for selection criteria. }
 \label{fig:scaling}
\end{figure}

Figure~\ref{fig:scaling}(a) compares five sizes from $L=32$ to $512$. In the two charge-$4e$ intervals, at $T=0.19$ and $1.2$, $S_{2e}$ levels off near 114 and 36, respectively. In the reentrant SC interval, at $T=0.56$, increasing $L$ from 128 to 512 raises $S_{2e}$ from $9.79\times10^3$ to $1.56\times10^5$. This nearly sixteenfold increase for a fourfold increase in $L$ is consistent with a small decay exponent on these sizes. At all three temperatures, $S_{4e}$ grows with $L$ and $S_{4e}/L^2$ decreases only weakly.

The real-space correlations show that charge-$2e$ coherence persists over distances comparable to the system size in the reentrant SC interval [Fig.~\ref{fig:scaling}(b)]. On the $L=512$ lattice, $G_{2e}(L/2)$ is consistent with zero at both outer temperatures. At $T=0.56$, $G_{2e}(L/2)\simeq0.593$. The charge-$4e$ correlations decay slowly at all three temperatures. These size and distance dependences support restored algebraic charge-$2e$ coherence without assuming a critical exponent for the recovery transition.

Figure~\ref{fig:scaling}(c) follows the two channels from $T=0.08$ to $6$. We define $R_{2me}(L,T)=S_{2me}(2L,T)/S_{2me}(L,T)$, with $L$ the smaller size. This ratio tends to $2^{2-\eta_{2me}}$ for algebraic correlations and to one for short-range correlations. Across $T_1$, $R_{2e}$ falls from near four toward one. It recovers to near four on heating through $T_2$ and falls again across $T_3$. The charge-$4e$ ratio stays large through all three boundaries and approaches one on the normal side of $T_4$. The scan identifies the sequence, with the coherence boundaries located separately by finite-size scaling. At $T=5$, for example, $R_{4e}$ is $1.43$, $1.01$, and $1.00$ for the size pairs $64/128$, $128/256$, and $256/512$, respectively.

To examine critical scaling at the recovery boundary, write $\phi_i=2\theta_i\bmod 2\pi$ for the phase of the charge-$4e$ composite field. For a local choice of its half-angle, $e^{i\theta_i}=\sigma_i e^{i\phi_i/2}$, where $\sigma_i=\pm1$ labels the two phase branches. If the critical Ising fluctuations factorize from the gapless Gaussian fluctuations of $\phi_i$~\cite{Shi2011,Serna2017}, the charge-$2e$ decay exponent is
\begin{equation}
 \eta_{2e}=\frac14+\frac{\eta_{4e}}4.
 \label{eq:isinggaussian}
\end{equation}
The first term is the Ising spin exponent. The half-angle contributes $\eta_{4e}/4$ because the Gaussian decay exponent of $e^{ia\phi}$ is proportional to $a^2$. This suggests the dimensionless combination
\begin{equation}
 X(L,T)=L^{1/4}\frac{S_{2e}(L)/L^2}{[S_{4e}(L)/L^2]^{1/4}}.
 \label{eq:scalingobservables}
\end{equation}
The prefactor $L^{1/4}$ incorporates the Ising spin exponent as an input. A common crossing therefore tests the factorized Ising--Gaussian scaling hypothesis. Its slope as a function of temperature probes the thermal exponent $1/\nu$.

The curves of $X$ cross and steepen with increasing $L$ [Fig.~\ref{fig:scaling}(d)]. Of the 45 late-cutoff points shown, 40 pass the equilibration and sampling checks. Fitting these points for $64\leq L\leq768$ in the variable $(T-T_2)L^{1/\nu}$ gives $T_2\simeq0.35434$ and $1/\nu\simeq1.03$. For this fit, conditional $95\%$ bootstrap intervals are $[0.35431,0.35438]$ and $[0.96,1.11]$, respectively. Here $\nu$ describes the divergence of the Ising-branch correlation length. Fits including $L=1024$ are reported as a size-sensitivity check in the SM. The fitted thermal exponent is compatible with the two-dimensional Ising value $1/\nu=1$. The real-space ratio $R_{\mathrm I}=\mathcal C_{\mathrm I}(L/2)/\mathcal C_{\mathrm I}(L/4)$, with $\mathcal C_{\mathrm I}(r)=G_{2e}(r)/[G_{4e}(r)]^{1/4}$, gives a consistent crossing on the same histories (SM). The two losses of charge-$2e$ coherence occur near $T_1\simeq0.118$ and $T_3\simeq0.798$.

At the high-temperature end of the sequence, we measure the helicity modulus $\rho$ with respect to twists of the charge-$2e$ phase $\theta$. With this convention, half-vortex unbinding at a BKT transition would give $\rho(T_4^-)/T_4=8/\pi$. Finite-size stiffness fits using this normalization are compatible with charge-$4e$ melting near $T_4\simeq4.44$ (SM).

\runinhead{Domain walls and half-vortex confinement.}
The recovery of charge-$2e$ coherence also has implications for half vortices. A half vortex terminates a domain wall across which the charge-$2e$ field switches branches~\cite{LeeGrinstein1985,Shi2011,Serna2017}. A positive wall tension confines a half-vortex pair linearly. When this tension vanishes, the logarithmic attraction can still bind the pair in the charge-$4e$ phase.

In the constrained limit, an Ising domain wall crosses antialigned bonds. Relative to an aligned bond, each has the bare free-energy cost
\begin{equation}
 \Delta f_{\rm bond}(T)=2J_{\rm I}(T)=2TK(T).
 \label{eq:wallcost}
\end{equation}
Selective thermal activation gives the aligned bond more entropy, so this local cost increases with temperature. The interfacial tension also reflects the entropy of wall shapes. It is positive for $K>K_c$ and vanishes for $K\leq K_c$~\cite{Liaw1999,Wu2006}. Thus the three crossings in Fig.~\ref{fig:mechanism} imply that this tension disappears, returns, and disappears again on heating.

We test this prediction at finite $J_2$ using two wall geometries. For an open wall, let $Z_{+-}(R,T;L)$ be the partition function with a half vortex and a half antivortex pinned to two triangular faces a distance $R$ apart along a lattice direction. The wall between the endpoints fluctuates. All other faces have zero charge-$4e$ vorticity, and global windings are summed over. At $J_2=3$ the separation free energy of this isolated pair, relative to $R_0=2$, is
\begin{equation}
 \Delta F_{1/2}(R,T;L)=-T\ln
 \frac{Z_{+-}(R,T;L)}{Z_{+-}(R_0,T;L)}.
 \label{eq:pairfreeenergy}
\end{equation}
At fixed $T$, we write the separation free energy as
\begin{equation}
 \begin{aligned}
 \Delta F_{1/2}(R,T;L)={}&\tau_{\rm pair}(R-R_0)+A\ln(R/R_0)\\
                         &+u_L(R)-u_L(R_0).
 \end{aligned}
 \label{eq:stringtension}
\end{equation}
The correction $u_L$ includes short-distance and finite-size effects and can be appreciable at $R_0=2$. For $1\ll R\ll L$, the slope of a linearly confining potential approaches the string tension $\tau_{\rm pair}>0$. For a logarithmic interaction it decays as $A/R$.

\begin{figure}[!htbp]
 \centering
 \includegraphics[width=\columnwidth]{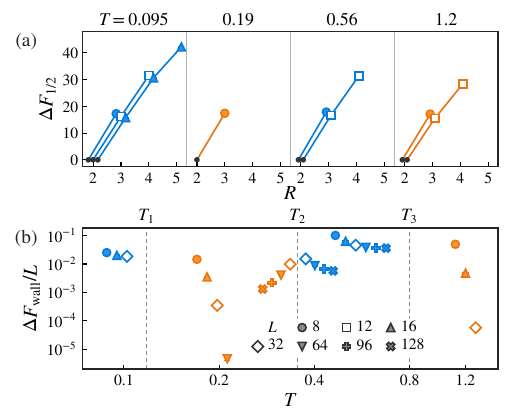}
 \caption[Half-vortex pairs and system-spanning domain walls]{Defect free energies at $J_2=3$ and $\varepsilon=0.235$. (a) Separation free energy of an isolated half-vortex pair relative to $R=2$. Lines connect measurements at fixed $L$ and $T$. (b) Excess free energy per unit length, $\Delta F_{\rm wall}/L$, of a system-spanning domain wall in the vortex-free, zero-winding sector of the charge-$4e$ field. The measurements at $T=0.34$ and $0.37$ bracket $T_2$ and span $L=32$ to $128$. Dashed lines mark $T_1$, $T_2$, and $T_3$. In both panels colors identify SC (blue) and charge-$4e$ (orange) temperatures. Symbols identify $L$ and are offset horizontally for clarity. Points pass the checks at both equilibration cutoffs. Bars give one standard error at the later cutoff.}
 \label{fig:confinement}
\end{figure}

Figure~\ref{fig:confinement}(a) shows the isolated-pair measurements in both SC and both charge-$4e$ intervals. The largest separation is $R=5$ in the low-temperature SC interval and $R=4$ in the reentrant SC interval. These distances do not distinguish an asymptotically linear potential from a logarithmic one.

The second geometry probes a domain wall that spans the periodic system. We impose antiperiodic (AP) or periodic (P) branch boundary conditions in the sector with zero charge-$4e$ vorticity and zero global winding. The excess free energy and the interfacial tension in this sector are
\begin{equation}
 \begin{aligned}
 \Delta F_{\rm wall}(L,T)&=-T\ln
 \frac{Z_{\rm AP}^{(0)}(L,T)}{Z_{\rm P}^{(0)}(L,T)},\\
 \tau_{\rm wall}^{(0)}(T)&=\lim_{L\to\infty}\frac{\Delta F_{\rm wall}(L,T)}{L}.
 \end{aligned}
 \label{eq:wallfreeenergy}
\end{equation}
Here $Z_{\rm AP}^{(0)}$ is the arithmetic mean over the two equivalent seam orientations. Figure~\ref{fig:confinement}(b) shows the cost per unit length, which decreases with size just below and just above $T_2$. The size dependence of the total excess free energy reverses across this boundary. Over the four sizes $L=32$, 64, 96, and 128, $\Delta F_{\rm wall}$ decreases monotonically from $0.3138(16)$ to $0.1674(11)$ at $T=0.34$. At $T=0.37$ it increases at each step, from $0.4839(22)$ to $0.7326(34)$. Deeper in the reentrant SC interval, at $T=0.56$, $\Delta F_{\rm wall}$ grows from $3.502(27)$ at $L=96$ to $4.619(24)$ at $L=128$. The corresponding costs per unit length, $0.03648(28)$ and $0.03608(19)$, agree within their uncertainties.

The near-constant $\Delta F_{\rm wall}/L$ at the two largest sizes is consistent with a linear wall cost at $T=0.56$. Determining $\tau_{\rm wall}^{(0)}$ requires a controlled large-$L$ extrapolation. The string tension $\tau_{\rm pair}$ probes a different defect ensemble and requires a separate large-distance limit.

\runinhead{Discussion and outlook.}
The charge-$4e$ phase plays two roles along the heating path. It emerges when Cooper-pair coherence first melts and provides the coherent background for its return. The entropy gained by aligned bonds can thus reverse one stage of superconducting melting at fixed microscopic couplings.

Exchanging the two charge-$2e$ branches leaves the charge-$4e$ field unchanged. A domain wall therefore probes the order that recovers within the coherent background. Its tension returns with Ising branch order in the constrained limit. At finite $J_2$, the growing cost of a system-spanning wall in the projected ensemble is a signature of the same recovery. At the higher temperature where charge-$4e$ coherence melts, the helicity modulus is consistent with half-vortex unbinding~\cite{KT1973,Nelson1977}.

Selective thermal activation can also restore order among three phase branches. With a third harmonic, the constrained model maps onto the three-state Potts model and supports reentrant branch order over an open parameter region (SM). Whether this mechanism produces reentrant charge-$6e$ superconductivity at finite $J_3$, or operates in a microscopic superconductor, remains open.

\begin{acknowledgments}
\runinhead{Acknowledgments.}
We thank Fei Teng for helpful discussions. This work was supported by the National Key Research and Development Program of China (Grant No.~2022YFA1404204), the National Natural Science Foundation of China (Grant No.~12274086), and the Quantum Science and Technology--National Science and Technology Major Project (Grant No.~2024ZD0300104).
\end{acknowledgments}

\pdfbookmark[1]{Data availability}{data-availability}
\runinhead{Data availability}
The data that support the findings of this article are not publicly available. The data are available from the authors upon reasonable request.

\makeatletter\immediate\write\@auxout{\string\citation{apsrev42Control}}\makeatother
\nocite{Weber1988,Hsieh2013,Swendsen1987,Onsager1944,WuGuo2012}

\bibliographystyle{apsrev4-2}
\bibliography{references}

\begin{thebibliography}{57}%
\makeatletter
\providecommand \@ifxundefined [1]{%
 \@ifx{#1\undefined}
}%
\providecommand \@ifnum [1]{%
 \ifnum #1\expandafter \@firstoftwo
 \else \expandafter \@secondoftwo
 \fi
}%
\providecommand \@ifx [1]{%
 \ifx #1\expandafter \@firstoftwo
 \else \expandafter \@secondoftwo
 \fi
}%
\providecommand \natexlab [1]{#1}%
\providecommand \enquote  [1]{``#1''}%
\providecommand \bibnamefont  [1]{#1}%
\providecommand \bibfnamefont [1]{#1}%
\providecommand \citenamefont [1]{#1}%
\providecommand \href@noop [0]{\@secondoftwo}%
\providecommand \href [0]{\begingroup \@sanitize@url \@href}%
\providecommand \@href[1]{\@@startlink{#1}\@@href}%
\providecommand \@@href[1]{\endgroup#1\@@endlink}%
\providecommand \@sanitize@url [0]{\catcode `\\12\catcode `\$12\catcode `\&12\catcode `\#12\catcode `\^12\catcode `\_12\catcode `\%12\relax}%
\providecommand \@@startlink[1]{}%
\providecommand \@@endlink[0]{}%
\providecommand \url  [0]{\begingroup\@sanitize@url \@url }%
\providecommand \@url [1]{\endgroup\@href {#1}{\urlprefix }}%
\providecommand \urlprefix  [0]{URL }%
\providecommand \Eprint [0]{\href }%
\providecommand \doibase [0]{https://doi.org/}%
\providecommand \selectlanguage [0]{\@gobble}%
\providecommand \bibinfo  [0]{\@secondoftwo}%
\providecommand \bibfield  [0]{\@secondoftwo}%
\providecommand \translation [1]{[#1]}%
\providecommand \BibitemOpen [0]{}%
\providecommand \bibitemStop [0]{}%
\providecommand \bibitemNoStop [0]{.\EOS\space}%
\providecommand \EOS [0]{\spacefactor3000\relax}%
\providecommand \BibitemShut  [1]{\csname bibitem#1\endcsname}%
\let\auto@bib@innerbib\@empty
\bibitem [{\citenamefont {Halperin}\ and\ \citenamefont {Nelson}(1978)}]{HalperinNelson1978}%
  \BibitemOpen
  \bibfield  {author} {\bibinfo {author} {\bibfnamefont {B.~I.}\ \bibnamefont {Halperin}}\ and\ \bibinfo {author} {\bibfnamefont {D.~R.}\ \bibnamefont {Nelson}},\ }\bibfield  {title} {\bibinfo {title} {Theory of two-dimensional melting},\ }\href {https://doi.org/10.1103/PhysRevLett.41.121} {\bibfield  {journal} {\bibinfo  {journal} {Phys. Rev. Lett.}\ }\textbf {\bibinfo {volume} {41}},\ \bibinfo {pages} {121} (\bibinfo {year} {1978})}\BibitemShut {NoStop}%
\bibitem [{\citenamefont {Nelson}\ and\ \citenamefont {Halperin}(1979)}]{NelsonHalperin1979}%
  \BibitemOpen
  \bibfield  {author} {\bibinfo {author} {\bibfnamefont {D.~R.}\ \bibnamefont {Nelson}}\ and\ \bibinfo {author} {\bibfnamefont {B.~I.}\ \bibnamefont {Halperin}},\ }\bibfield  {title} {\bibinfo {title} {Dislocation-mediated melting in two dimensions},\ }\href {https://doi.org/10.1103/PhysRevB.19.2457} {\bibfield  {journal} {\bibinfo  {journal} {Phys. Rev. B}\ }\textbf {\bibinfo {volume} {19}},\ \bibinfo {pages} {2457} (\bibinfo {year} {1979})}\BibitemShut {NoStop}%
\bibitem [{\citenamefont {Young}(1979)}]{Young1979}%
  \BibitemOpen
  \bibfield  {author} {\bibinfo {author} {\bibfnamefont {A.~P.}\ \bibnamefont {Young}},\ }\bibfield  {title} {\bibinfo {title} {Melting and the vector {Coulomb} gas in two dimensions},\ }\href {https://doi.org/10.1103/PhysRevB.19.1855} {\bibfield  {journal} {\bibinfo  {journal} {Phys. Rev. B}\ }\textbf {\bibinfo {volume} {19}},\ \bibinfo {pages} {1855} (\bibinfo {year} {1979})}\BibitemShut {NoStop}%
\bibitem [{\citenamefont {Chandra}\ \emph {et~al.}(1990)\citenamefont {Chandra}, \citenamefont {Coleman},\ and\ \citenamefont {Larkin}}]{Chandra1990}%
  \BibitemOpen
  \bibfield  {author} {\bibinfo {author} {\bibfnamefont {P.}~\bibnamefont {Chandra}}, \bibinfo {author} {\bibfnamefont {P.}~\bibnamefont {Coleman}},\ and\ \bibinfo {author} {\bibfnamefont {A.~I.}\ \bibnamefont {Larkin}},\ }\bibfield  {title} {\bibinfo {title} {{Ising} transition in frustrated {Heisenberg} models},\ }\href {https://doi.org/10.1103/PhysRevLett.64.88} {\bibfield  {journal} {\bibinfo  {journal} {Phys. Rev. Lett.}\ }\textbf {\bibinfo {volume} {64}},\ \bibinfo {pages} {88} (\bibinfo {year} {1990})}\BibitemShut {NoStop}%
\bibitem [{\citenamefont {Kivelson}\ \emph {et~al.}(1998)\citenamefont {Kivelson}, \citenamefont {Fradkin},\ and\ \citenamefont {Emery}}]{Kivelson1998}%
  \BibitemOpen
  \bibfield  {author} {\bibinfo {author} {\bibfnamefont {S.~A.}\ \bibnamefont {Kivelson}}, \bibinfo {author} {\bibfnamefont {E.}~\bibnamefont {Fradkin}},\ and\ \bibinfo {author} {\bibfnamefont {V.~J.}\ \bibnamefont {Emery}},\ }\bibfield  {title} {\bibinfo {title} {Electronic liquid-crystal phases of a doped {Mott} insulator},\ }\href {https://doi.org/10.1038/31177} {\bibfield  {journal} {\bibinfo  {journal} {Nature}\ }\textbf {\bibinfo {volume} {393}},\ \bibinfo {pages} {550} (\bibinfo {year} {1998})}\BibitemShut {NoStop}%
\bibitem [{\citenamefont {Fernandes}\ \emph {et~al.}(2012)\citenamefont {Fernandes}, \citenamefont {Chubukov}, \citenamefont {Knolle}, \citenamefont {Eremin},\ and\ \citenamefont {Schmalian}}]{Fernandes2012}%
  \BibitemOpen
  \bibfield  {author} {\bibinfo {author} {\bibfnamefont {R.~M.}\ \bibnamefont {Fernandes}}, \bibinfo {author} {\bibfnamefont {A.~V.}\ \bibnamefont {Chubukov}}, \bibinfo {author} {\bibfnamefont {J.}~\bibnamefont {Knolle}}, \bibinfo {author} {\bibfnamefont {I.}~\bibnamefont {Eremin}},\ and\ \bibinfo {author} {\bibfnamefont {J.}~\bibnamefont {Schmalian}},\ }\bibfield  {title} {\bibinfo {title} {Preemptive nematic order, pseudogap, and orbital order in the iron pnictides},\ }\href {https://doi.org/10.1103/PhysRevB.85.024534} {\bibfield  {journal} {\bibinfo  {journal} {Phys. Rev. B}\ }\textbf {\bibinfo {volume} {85}},\ \bibinfo {pages} {024534} (\bibinfo {year} {2012})}\BibitemShut {NoStop}%
\bibitem [{\citenamefont {Fernandes}\ \emph {et~al.}(2019)\citenamefont {Fernandes}, \citenamefont {Orth},\ and\ \citenamefont {Schmalian}}]{FernandesOrthSchmalian2019}%
  \BibitemOpen
  \bibfield  {author} {\bibinfo {author} {\bibfnamefont {R.~M.}\ \bibnamefont {Fernandes}}, \bibinfo {author} {\bibfnamefont {P.~P.}\ \bibnamefont {Orth}},\ and\ \bibinfo {author} {\bibfnamefont {J.}~\bibnamefont {Schmalian}},\ }\bibfield  {title} {\bibinfo {title} {Intertwined vestigial order in quantum materials: Nematicity and beyond},\ }\href {https://doi.org/10.1146/annurev-conmatphys-031218-013200} {\bibfield  {journal} {\bibinfo  {journal} {Annu. Rev. Condens. Matter Phys.}\ }\textbf {\bibinfo {volume} {10}},\ \bibinfo {pages} {133} (\bibinfo {year} {2019})}\BibitemShut {NoStop}%
\bibitem [{\citenamefont {Agterberg}\ and\ \citenamefont {Tsunetsugu}(2008)}]{Agterberg2008}%
  \BibitemOpen
  \bibfield  {author} {\bibinfo {author} {\bibfnamefont {D.~F.}\ \bibnamefont {Agterberg}}\ and\ \bibinfo {author} {\bibfnamefont {H.}~\bibnamefont {Tsunetsugu}},\ }\bibfield  {title} {\bibinfo {title} {Dislocations and vortices in pair-density-wave superconductors},\ }\href {https://doi.org/10.1038/nphys999} {\bibfield  {journal} {\bibinfo  {journal} {Nat. Phys.}\ }\textbf {\bibinfo {volume} {4}},\ \bibinfo {pages} {639} (\bibinfo {year} {2008})}\BibitemShut {NoStop}%
\bibitem [{\citenamefont {Berg}\ \emph {et~al.}(2009)\citenamefont {Berg}, \citenamefont {Fradkin},\ and\ \citenamefont {Kivelson}}]{Berg2009}%
  \BibitemOpen
  \bibfield  {author} {\bibinfo {author} {\bibfnamefont {E.}~\bibnamefont {Berg}}, \bibinfo {author} {\bibfnamefont {E.}~\bibnamefont {Fradkin}},\ and\ \bibinfo {author} {\bibfnamefont {S.~A.}\ \bibnamefont {Kivelson}},\ }\bibfield  {title} {\bibinfo {title} {Charge-{$4e$} superconductivity from pair-density-wave order in certain high-temperature superconductors},\ }\href {https://doi.org/10.1038/nphys1389} {\bibfield  {journal} {\bibinfo  {journal} {Nat. Phys.}\ }\textbf {\bibinfo {volume} {5}},\ \bibinfo {pages} {830} (\bibinfo {year} {2009})}\BibitemShut {NoStop}%
\bibitem [{\citenamefont {Agterberg}\ \emph {et~al.}(2020)\citenamefont {Agterberg}, \citenamefont {Davis}, \citenamefont {Edkins}, \citenamefont {Fradkin}, \citenamefont {Van~Harlingen}, \citenamefont {Kivelson}, \citenamefont {Lee}, \citenamefont {Radzihovsky}, \citenamefont {Tranquada},\ and\ \citenamefont {Wang}}]{Agterberg2020}%
  \BibitemOpen
  \bibfield  {author} {\bibinfo {author} {\bibfnamefont {D.~F.}\ \bibnamefont {Agterberg}}, \bibinfo {author} {\bibfnamefont {J.~C.~S.}\ \bibnamefont {Davis}}, \bibinfo {author} {\bibfnamefont {S.~D.}\ \bibnamefont {Edkins}}, \bibinfo {author} {\bibfnamefont {E.}~\bibnamefont {Fradkin}}, \bibinfo {author} {\bibfnamefont {D.~J.}\ \bibnamefont {Van~Harlingen}}, \bibinfo {author} {\bibfnamefont {S.~A.}\ \bibnamefont {Kivelson}}, \bibinfo {author} {\bibfnamefont {P.~A.}\ \bibnamefont {Lee}}, \bibinfo {author} {\bibfnamefont {L.}~\bibnamefont {Radzihovsky}}, \bibinfo {author} {\bibfnamefont {J.~M.}\ \bibnamefont {Tranquada}},\ and\ \bibinfo {author} {\bibfnamefont {Y.}~\bibnamefont {Wang}},\ }\bibfield  {title} {\bibinfo {title} {The physics of pair-density waves: Cuprate superconductors and beyond},\ }\href {https://doi.org/10.1146/annurev-conmatphys-031119-050711} {\bibfield  {journal} {\bibinfo  {journal} {Annu. Rev. Condens. Matter Phys.}\ }\textbf {\bibinfo {volume} {11}},\ \bibinfo {pages} {231} (\bibinfo
  {year} {2020})}\BibitemShut {NoStop}%
\bibitem [{\citenamefont {Wu}\ and\ \citenamefont {Wang}(2024)}]{WuWang2024}%
  \BibitemOpen
  \bibfield  {author} {\bibinfo {author} {\bibfnamefont {Y.-M.}\ \bibnamefont {Wu}}\ and\ \bibinfo {author} {\bibfnamefont {Y.}~\bibnamefont {Wang}},\ }\bibfield  {title} {\bibinfo {title} {{$d$}-wave charge-{$4e$} superconductivity from fluctuating pair density waves},\ }\href {https://doi.org/10.1038/s41535-024-00674-y} {\bibfield  {journal} {\bibinfo  {journal} {npj Quantum Mater.}\ }\textbf {\bibinfo {volume} {9}},\ \bibinfo {pages} {66} (\bibinfo {year} {2024})}\BibitemShut {NoStop}%
\bibitem [{\citenamefont {Huecker}\ and\ \citenamefont {Wang}(2026{\natexlab{a}})}]{HueckerWang2026}%
  \BibitemOpen
  \bibfield  {author} {\bibinfo {author} {\bibfnamefont {E.}~\bibnamefont {Huecker}}\ and\ \bibinfo {author} {\bibfnamefont {Y.}~\bibnamefont {Wang}},\ }\bibfield  {title} {\bibinfo {title} {Vestigial {$d$}-wave charge-{$4e$} superconductivity from bidirectional pair density waves},\ }\href {https://doi.org/10.1103/7lbj-tzw3} {\bibfield  {journal} {\bibinfo  {journal} {Phys. Rev. B}\ }\textbf {\bibinfo {volume} {113}},\ \bibinfo {pages} {045121} (\bibinfo {year} {2026}{\natexlab{a}})}\BibitemShut {NoStop}%
\bibitem [{\citenamefont {Huecker}\ and\ \citenamefont {Wang}(2026{\natexlab{b}})}]{HueckerWangErratum2026}%
  \BibitemOpen
  \bibfield  {author} {\bibinfo {author} {\bibfnamefont {E.}~\bibnamefont {Huecker}}\ and\ \bibinfo {author} {\bibfnamefont {Y.}~\bibnamefont {Wang}},\ }\bibfield  {title} {\bibinfo {title} {Erratum: Vestigial {$d$}-wave charge-{$4e$} superconductivity from bidirectional pair density waves [{Phys. Rev. B} 113, 045121 (2026)]},\ }\href {https://doi.org/10.1103/j9yw-3rbf} {\bibfield  {journal} {\bibinfo  {journal} {Phys. Rev. B}\ }\textbf {\bibinfo {volume} {114}},\ \bibinfo {pages} {229901} (\bibinfo {year} {2026}{\natexlab{b}})}\BibitemShut {NoStop}%
\bibitem [{\citenamefont {Radzihovsky}\ and\ \citenamefont {Vishwanath}(2009)}]{RadzihovskyVishwanath2009}%
  \BibitemOpen
  \bibfield  {author} {\bibinfo {author} {\bibfnamefont {L.}~\bibnamefont {Radzihovsky}}\ and\ \bibinfo {author} {\bibfnamefont {A.}~\bibnamefont {Vishwanath}},\ }\bibfield  {title} {\bibinfo {title} {Quantum liquid crystals in an imbalanced {Fermi} gas: Fluctuations and fractional vortices in {Larkin-Ovchinnikov} states},\ }\href {https://doi.org/10.1103/PhysRevLett.103.010404} {\bibfield  {journal} {\bibinfo  {journal} {Phys. Rev. Lett.}\ }\textbf {\bibinfo {volume} {103}},\ \bibinfo {pages} {010404} (\bibinfo {year} {2009})}\BibitemShut {NoStop}%
\bibitem [{\citenamefont {Radzihovsky}(2011)}]{Radzihovsky2011}%
  \BibitemOpen
  \bibfield  {author} {\bibinfo {author} {\bibfnamefont {L.}~\bibnamefont {Radzihovsky}},\ }\bibfield  {title} {\bibinfo {title} {Fluctuations and phase transitions in {Larkin-Ovchinnikov} liquid-crystal states of a population-imbalanced resonant {Fermi} gas},\ }\href {https://doi.org/10.1103/PhysRevA.84.023611} {\bibfield  {journal} {\bibinfo  {journal} {Phys. Rev. A}\ }\textbf {\bibinfo {volume} {84}},\ \bibinfo {pages} {023611} (\bibinfo {year} {2011})}\BibitemShut {NoStop}%
\bibitem [{\citenamefont {Fernandes}\ and\ \citenamefont {Fu}(2021)}]{FernandesFu2021}%
  \BibitemOpen
  \bibfield  {author} {\bibinfo {author} {\bibfnamefont {R.~M.}\ \bibnamefont {Fernandes}}\ and\ \bibinfo {author} {\bibfnamefont {L.}~\bibnamefont {Fu}},\ }\bibfield  {title} {\bibinfo {title} {Charge-{$4e$} superconductivity from multicomponent nematic pairing: Application to twisted bilayer graphene},\ }\href {https://doi.org/10.1103/PhysRevLett.127.047001} {\bibfield  {journal} {\bibinfo  {journal} {Phys. Rev. Lett.}\ }\textbf {\bibinfo {volume} {127}},\ \bibinfo {pages} {047001} (\bibinfo {year} {2021})}\BibitemShut {NoStop}%
\bibitem [{\citenamefont {Jian}\ \emph {et~al.}(2021)\citenamefont {Jian}, \citenamefont {Huang},\ and\ \citenamefont {Yao}}]{Jian2021}%
  \BibitemOpen
  \bibfield  {author} {\bibinfo {author} {\bibfnamefont {S.-K.}\ \bibnamefont {Jian}}, \bibinfo {author} {\bibfnamefont {Y.}~\bibnamefont {Huang}},\ and\ \bibinfo {author} {\bibfnamefont {H.}~\bibnamefont {Yao}},\ }\bibfield  {title} {\bibinfo {title} {Charge-{$4e$} superconductivity from nematic superconductors in two and three dimensions},\ }\href {https://doi.org/10.1103/PhysRevLett.127.227001} {\bibfield  {journal} {\bibinfo  {journal} {Phys. Rev. Lett.}\ }\textbf {\bibinfo {volume} {127}},\ \bibinfo {pages} {227001} (\bibinfo {year} {2021})}\BibitemShut {NoStop}%
\bibitem [{\citenamefont {Hecker}\ \emph {et~al.}(2023)\citenamefont {Hecker}, \citenamefont {Willa}, \citenamefont {Schmalian},\ and\ \citenamefont {Fernandes}}]{Hecker2023}%
  \BibitemOpen
  \bibfield  {author} {\bibinfo {author} {\bibfnamefont {M.}~\bibnamefont {Hecker}}, \bibinfo {author} {\bibfnamefont {R.}~\bibnamefont {Willa}}, \bibinfo {author} {\bibfnamefont {J.}~\bibnamefont {Schmalian}},\ and\ \bibinfo {author} {\bibfnamefont {R.~M.}\ \bibnamefont {Fernandes}},\ }\bibfield  {title} {\bibinfo {title} {Cascade of vestigial orders in two-component superconductors: Nematic, ferromagnetic, {$s$}-wave charge-{$4e$}, and {$d$}-wave charge-{$4e$} states},\ }\href {https://doi.org/10.1103/PhysRevB.107.224503} {\bibfield  {journal} {\bibinfo  {journal} {Phys. Rev. B}\ }\textbf {\bibinfo {volume} {107}},\ \bibinfo {pages} {224503} (\bibinfo {year} {2023})}\BibitemShut {NoStop}%
\bibitem [{\citenamefont {Hecker}\ and\ \citenamefont {Fernandes}(2024)}]{HeckerFernandes2024}%
  \BibitemOpen
  \bibfield  {author} {\bibinfo {author} {\bibfnamefont {M.}~\bibnamefont {Hecker}}\ and\ \bibinfo {author} {\bibfnamefont {R.~M.}\ \bibnamefont {Fernandes}},\ }\bibfield  {title} {\bibinfo {title} {Local condensation of charge-{$4e$} superconductivity at a nematic domain wall},\ }\href {https://doi.org/10.1103/PhysRevB.109.134514} {\bibfield  {journal} {\bibinfo  {journal} {Phys. Rev. B}\ }\textbf {\bibinfo {volume} {109}},\ \bibinfo {pages} {134514} (\bibinfo {year} {2024})}\BibitemShut {NoStop}%
\bibitem [{\citenamefont {Liu}\ \emph {et~al.}(2024)\citenamefont {Liu}, \citenamefont {Zhou},\ and\ \citenamefont {Yang}}]{Liu2024}%
  \BibitemOpen
  \bibfield  {author} {\bibinfo {author} {\bibfnamefont {Y.-B.}\ \bibnamefont {Liu}}, \bibinfo {author} {\bibfnamefont {J.}~\bibnamefont {Zhou}},\ and\ \bibinfo {author} {\bibfnamefont {F.}~\bibnamefont {Yang}},\ }\bibfield  {title} {\bibinfo {title} {Nematic superconductivity and its critical vestigial phases in the quasicrystal},\ }\href {https://doi.org/10.1103/PhysRevLett.133.136002} {\bibfield  {journal} {\bibinfo  {journal} {Phys. Rev. Lett.}\ }\textbf {\bibinfo {volume} {133}},\ \bibinfo {pages} {136002} (\bibinfo {year} {2024})}\BibitemShut {NoStop}%
\bibitem [{\citenamefont {Zou}\ \emph {et~al.}(2026)\citenamefont {Zou}, \citenamefont {Wan},\ and\ \citenamefont {Yao}}]{Zou2026}%
  \BibitemOpen
  \bibfield  {author} {\bibinfo {author} {\bibfnamefont {X.}~\bibnamefont {Zou}}, \bibinfo {author} {\bibfnamefont {Z.-Q.}\ \bibnamefont {Wan}},\ and\ \bibinfo {author} {\bibfnamefont {H.}~\bibnamefont {Yao}},\ }\bibfield  {title} {\bibinfo {title} {Emergence of charge-{$4e$} superconductivity from {2D} nematic superconductors},\ }\href {https://doi.org/10.1103/j7nh-jzdd} {\bibfield  {journal} {\bibinfo  {journal} {Phys. Rev. Lett.}\ }\textbf {\bibinfo {volume} {137}},\ \bibinfo {pages} {066505} (\bibinfo {year} {2026})}\BibitemShut {NoStop}%
\bibitem [{\citenamefont {Maccari}\ \emph {et~al.}(2026)\citenamefont {Maccari}, \citenamefont {Babaev},\ and\ \citenamefont {Carlstr{\"o}m}}]{Maccari2026}%
  \BibitemOpen
  \bibfield  {author} {\bibinfo {author} {\bibfnamefont {I.}~\bibnamefont {Maccari}}, \bibinfo {author} {\bibfnamefont {E.}~\bibnamefont {Babaev}},\ and\ \bibinfo {author} {\bibfnamefont {J.}~\bibnamefont {Carlstr{\"o}m}},\ }\bibfield  {title} {\bibinfo {title} {Revisiting vestigial order in nematic superconductors: Gauge-field mechanisms and model constraints},\ }\href {https://doi.org/10.1103/gccc-rfw4} {\bibfield  {journal} {\bibinfo  {journal} {Phys. Rev. B}\ }\textbf {\bibinfo {volume} {113}},\ \bibinfo {pages} {014501} (\bibinfo {year} {2026})}\BibitemShut {NoStop}%
\bibitem [{\citenamefont {Dou{\c c}ot}\ and\ \citenamefont {Vidal}(2002)}]{DoucotVidal2002}%
  \BibitemOpen
  \bibfield  {author} {\bibinfo {author} {\bibfnamefont {B.}~\bibnamefont {Dou{\c c}ot}}\ and\ \bibinfo {author} {\bibfnamefont {J.}~\bibnamefont {Vidal}},\ }\bibfield  {title} {\bibinfo {title} {Pairing of {Cooper} pairs in a fully frustrated {Josephson}-junction chain},\ }\href {https://doi.org/10.1103/PhysRevLett.88.227005} {\bibfield  {journal} {\bibinfo  {journal} {Phys. Rev. Lett.}\ }\textbf {\bibinfo {volume} {88}},\ \bibinfo {pages} {227005} (\bibinfo {year} {2002})}\BibitemShut {NoStop}%
\bibitem [{\citenamefont {Song}\ and\ \citenamefont {Zhang}(2022)}]{SongZhang2022}%
  \BibitemOpen
  \bibfield  {author} {\bibinfo {author} {\bibfnamefont {F.-F.}\ \bibnamefont {Song}}\ and\ \bibinfo {author} {\bibfnamefont {G.-M.}\ \bibnamefont {Zhang}},\ }\bibfield  {title} {\bibinfo {title} {Phase coherence of pairs of {Cooper} pairs as quasi-long-range order of half-vortex pairs in a two-dimensional bilayer system},\ }\href {https://doi.org/10.1103/PhysRevLett.128.195301} {\bibfield  {journal} {\bibinfo  {journal} {Phys. Rev. Lett.}\ }\textbf {\bibinfo {volume} {128}},\ \bibinfo {pages} {195301} (\bibinfo {year} {2022})}\BibitemShut {NoStop}%
\bibitem [{\citenamefont {Liu}\ \emph {et~al.}(2023)\citenamefont {Liu}, \citenamefont {Zhou}, \citenamefont {Wu},\ and\ \citenamefont {Yang}}]{Liu2023}%
  \BibitemOpen
  \bibfield  {author} {\bibinfo {author} {\bibfnamefont {Y.-B.}\ \bibnamefont {Liu}}, \bibinfo {author} {\bibfnamefont {J.}~\bibnamefont {Zhou}}, \bibinfo {author} {\bibfnamefont {C.}~\bibnamefont {Wu}},\ and\ \bibinfo {author} {\bibfnamefont {F.}~\bibnamefont {Yang}},\ }\bibfield  {title} {\bibinfo {title} {Charge-{$4e$} superconductivity and chiral metal in {$45^\circ$}-twisted bilayer cuprates and related bilayers},\ }\href {https://doi.org/10.1038/s41467-023-43782-2} {\bibfield  {journal} {\bibinfo  {journal} {Nat. Commun.}\ }\textbf {\bibinfo {volume} {14}},\ \bibinfo {pages} {7926} (\bibinfo {year} {2023})}\BibitemShut {NoStop}%
\bibitem [{\citenamefont {Ge}\ \emph {et~al.}(2024)\citenamefont {Ge}, \citenamefont {Wang}, \citenamefont {Xing}, \citenamefont {Yin}, \citenamefont {Wang}, \citenamefont {Shen}, \citenamefont {Lei}, \citenamefont {Wang},\ and\ \citenamefont {Wang}}]{Ge2024}%
  \BibitemOpen
  \bibfield  {author} {\bibinfo {author} {\bibfnamefont {J.}~\bibnamefont {Ge}}, \bibinfo {author} {\bibfnamefont {P.}~\bibnamefont {Wang}}, \bibinfo {author} {\bibfnamefont {Y.}~\bibnamefont {Xing}}, \bibinfo {author} {\bibfnamefont {Q.}~\bibnamefont {Yin}}, \bibinfo {author} {\bibfnamefont {A.}~\bibnamefont {Wang}}, \bibinfo {author} {\bibfnamefont {J.}~\bibnamefont {Shen}}, \bibinfo {author} {\bibfnamefont {H.}~\bibnamefont {Lei}}, \bibinfo {author} {\bibfnamefont {Z.}~\bibnamefont {Wang}},\ and\ \bibinfo {author} {\bibfnamefont {J.}~\bibnamefont {Wang}},\ }\bibfield  {title} {\bibinfo {title} {Charge-$4e$ and charge-$6e$ flux quantization and higher charge superconductivity in kagome superconductor ring devices},\ }\href {https://doi.org/10.1103/PhysRevX.14.021025} {\bibfield  {journal} {\bibinfo  {journal} {Phys. Rev. X}\ }\textbf {\bibinfo {volume} {14}},\ \bibinfo {pages} {021025} (\bibinfo {year} {2024})}\BibitemShut {NoStop}%
\bibitem [{\citenamefont {Gladchenko}\ \emph {et~al.}(2009)\citenamefont {Gladchenko}, \citenamefont {Olaya}, \citenamefont {Dupont-Ferrier}, \citenamefont {Dou{\c{c}}ot}, \citenamefont {Ioffe},\ and\ \citenamefont {Gershenson}}]{Gladchenko2009}%
  \BibitemOpen
  \bibfield  {author} {\bibinfo {author} {\bibfnamefont {S.}~\bibnamefont {Gladchenko}}, \bibinfo {author} {\bibfnamefont {D.}~\bibnamefont {Olaya}}, \bibinfo {author} {\bibfnamefont {E.}~\bibnamefont {Dupont-Ferrier}}, \bibinfo {author} {\bibfnamefont {B.}~\bibnamefont {Dou{\c{c}}ot}}, \bibinfo {author} {\bibfnamefont {L.~B.}\ \bibnamefont {Ioffe}},\ and\ \bibinfo {author} {\bibfnamefont {M.~E.}\ \bibnamefont {Gershenson}},\ }\bibfield  {title} {\bibinfo {title} {Superconducting nanocircuits for topologically protected qubits},\ }\href {https://doi.org/10.1038/nphys1151} {\bibfield  {journal} {\bibinfo  {journal} {Nat. Phys.}\ }\textbf {\bibinfo {volume} {5}},\ \bibinfo {pages} {48} (\bibinfo {year} {2009})}\BibitemShut {NoStop}%
\bibitem [{\citenamefont {Huang}\ \emph {et~al.}(2022)\citenamefont {Huang}, \citenamefont {Ronen}, \citenamefont {M{\'e}lin}, \citenamefont {Feinberg}, \citenamefont {Watanabe}, \citenamefont {Taniguchi},\ and\ \citenamefont {Kim}}]{Huang2022}%
  \BibitemOpen
  \bibfield  {author} {\bibinfo {author} {\bibfnamefont {K.-F.}\ \bibnamefont {Huang}}, \bibinfo {author} {\bibfnamefont {Y.}~\bibnamefont {Ronen}}, \bibinfo {author} {\bibfnamefont {R.}~\bibnamefont {M{\'e}lin}}, \bibinfo {author} {\bibfnamefont {D.}~\bibnamefont {Feinberg}}, \bibinfo {author} {\bibfnamefont {K.}~\bibnamefont {Watanabe}}, \bibinfo {author} {\bibfnamefont {T.}~\bibnamefont {Taniguchi}},\ and\ \bibinfo {author} {\bibfnamefont {P.}~\bibnamefont {Kim}},\ }\bibfield  {title} {\bibinfo {title} {Evidence for $4e$ charge of {Cooper} quartets in a biased multi-terminal graphene-based {Josephson} junction},\ }\href {https://doi.org/10.1038/s41467-022-30732-7} {\bibfield  {journal} {\bibinfo  {journal} {Nat. Commun.}\ }\textbf {\bibinfo {volume} {13}},\ \bibinfo {pages} {3032} (\bibinfo {year} {2022})}\BibitemShut {NoStop}%
\bibitem [{\citenamefont {Ciaccia}\ \emph {et~al.}(2024)\citenamefont {Ciaccia}, \citenamefont {Haller}, \citenamefont {Drachmann}, \citenamefont {Lindemann}, \citenamefont {Manfra}, \citenamefont {Schrade},\ and\ \citenamefont {Sch{\"o}nenberger}}]{Ciaccia2024}%
  \BibitemOpen
  \bibfield  {author} {\bibinfo {author} {\bibfnamefont {C.}~\bibnamefont {Ciaccia}}, \bibinfo {author} {\bibfnamefont {R.}~\bibnamefont {Haller}}, \bibinfo {author} {\bibfnamefont {A.~C.~C.}\ \bibnamefont {Drachmann}}, \bibinfo {author} {\bibfnamefont {T.}~\bibnamefont {Lindemann}}, \bibinfo {author} {\bibfnamefont {M.~J.}\ \bibnamefont {Manfra}}, \bibinfo {author} {\bibfnamefont {C.}~\bibnamefont {Schrade}},\ and\ \bibinfo {author} {\bibfnamefont {C.}~\bibnamefont {Sch{\"o}nenberger}},\ }\bibfield  {title} {\bibinfo {title} {Charge-$4e$ supercurrent in a two-dimensional {InAs-Al} superconductor-semiconductor heterostructure},\ }\href {https://doi.org/10.1038/s42005-024-01531-x} {\bibfield  {journal} {\bibinfo  {journal} {Commun. Phys.}\ }\textbf {\bibinfo {volume} {7}},\ \bibinfo {pages} {41} (\bibinfo {year} {2024})}\BibitemShut {NoStop}%
\bibitem [{\citenamefont {Leblanc}\ \emph {et~al.}(2024)\citenamefont {Leblanc}, \citenamefont {Tangchingchai}, \citenamefont {Momtaz}, \citenamefont {Kiyooka}, \citenamefont {Hartmann}, \citenamefont {Fernandez-Bada}, \citenamefont {Scher{\"u}bl}, \citenamefont {Brun}, \citenamefont {Schmitt}, \citenamefont {Zihlmann}, \citenamefont {Maurand}, \citenamefont {Dumur}, \citenamefont {{De Franceschi}},\ and\ \citenamefont {Lefloch}}]{Leblanc2024}%
  \BibitemOpen
  \bibfield  {author} {\bibinfo {author} {\bibfnamefont {A.}~\bibnamefont {Leblanc}}, \bibinfo {author} {\bibfnamefont {C.}~\bibnamefont {Tangchingchai}}, \bibinfo {author} {\bibfnamefont {Z.~S.}\ \bibnamefont {Momtaz}}, \bibinfo {author} {\bibfnamefont {E.}~\bibnamefont {Kiyooka}}, \bibinfo {author} {\bibfnamefont {J.-M.}\ \bibnamefont {Hartmann}}, \bibinfo {author} {\bibfnamefont {G.~T.}\ \bibnamefont {Fernandez-Bada}}, \bibinfo {author} {\bibfnamefont {Z.}~\bibnamefont {Scher{\"u}bl}}, \bibinfo {author} {\bibfnamefont {B.}~\bibnamefont {Brun}}, \bibinfo {author} {\bibfnamefont {V.}~\bibnamefont {Schmitt}}, \bibinfo {author} {\bibfnamefont {S.}~\bibnamefont {Zihlmann}}, \bibinfo {author} {\bibfnamefont {R.}~\bibnamefont {Maurand}}, \bibinfo {author} {\bibfnamefont {{\'E}.}~\bibnamefont {Dumur}}, \bibinfo {author} {\bibfnamefont {S.}~\bibnamefont {{De Franceschi}}},\ and\ \bibinfo {author} {\bibfnamefont {F.}~\bibnamefont {Lefloch}},\ }\bibfield  {title} {\bibinfo {title} {From nonreciprocal to charge-$4e$
  supercurrent in {Ge}-based {Josephson} devices with tunable harmonic content},\ }\href {https://doi.org/10.1103/PhysRevResearch.6.033281} {\bibfield  {journal} {\bibinfo  {journal} {Phys. Rev. Research}\ }\textbf {\bibinfo {volume} {6}},\ \bibinfo {pages} {033281} (\bibinfo {year} {2024})}\BibitemShut {NoStop}%
\bibitem [{\citenamefont {Berezinskii}(1971)}]{Berezinskii1971}%
  \BibitemOpen
  \bibfield  {author} {\bibinfo {author} {\bibfnamefont {V.~L.}\ \bibnamefont {Berezinskii}},\ }\bibfield  {title} {\bibinfo {title} {Destruction of long-range order in one-dimensional and two-dimensional systems having a continuous symmetry group. {I}. classical systems},\ }\href {https://jetp.ras.ru/cgi-bin/dn/e_032_03_0493.pdf} {\bibfield  {journal} {\bibinfo  {journal} {Sov. Phys. JETP}\ }\textbf {\bibinfo {volume} {32}},\ \bibinfo {pages} {493} (\bibinfo {year} {1971})}\BibitemShut {NoStop}%
\bibitem [{\citenamefont {Kosterlitz}\ and\ \citenamefont {Thouless}(1973)}]{KT1973}%
  \BibitemOpen
  \bibfield  {author} {\bibinfo {author} {\bibfnamefont {J.~M.}\ \bibnamefont {Kosterlitz}}\ and\ \bibinfo {author} {\bibfnamefont {D.~J.}\ \bibnamefont {Thouless}},\ }\bibfield  {title} {\bibinfo {title} {Ordering, metastability and phase transitions in two-dimensional systems},\ }\href {https://doi.org/10.1088/0022-3719/6/7/010} {\bibfield  {journal} {\bibinfo  {journal} {J. Phys. C: Solid State Phys.}\ }\textbf {\bibinfo {volume} {6}},\ \bibinfo {pages} {1181} (\bibinfo {year} {1973})}\BibitemShut {NoStop}%
\bibitem [{\citenamefont {Kosterlitz}(1974)}]{Kosterlitz1974}%
  \BibitemOpen
  \bibfield  {author} {\bibinfo {author} {\bibfnamefont {J.~M.}\ \bibnamefont {Kosterlitz}},\ }\bibfield  {title} {\bibinfo {title} {The critical properties of the two-dimensional {$XY$} model},\ }\href {https://doi.org/10.1088/0022-3719/7/6/005} {\bibfield  {journal} {\bibinfo  {journal} {J. Phys. C: Solid State Phys.}\ }\textbf {\bibinfo {volume} {7}},\ \bibinfo {pages} {1046} (\bibinfo {year} {1974})}\BibitemShut {NoStop}%
\bibitem [{\citenamefont {Lee}\ and\ \citenamefont {Grinstein}(1985)}]{LeeGrinstein1985}%
  \BibitemOpen
  \bibfield  {author} {\bibinfo {author} {\bibfnamefont {D.~H.}\ \bibnamefont {Lee}}\ and\ \bibinfo {author} {\bibfnamefont {G.}~\bibnamefont {Grinstein}},\ }\bibfield  {title} {\bibinfo {title} {Strings in two-dimensional classical {$XY$} models},\ }\href {https://doi.org/10.1103/PhysRevLett.55.541} {\bibfield  {journal} {\bibinfo  {journal} {Phys. Rev. Lett.}\ }\textbf {\bibinfo {volume} {55}},\ \bibinfo {pages} {541} (\bibinfo {year} {1985})}\BibitemShut {NoStop}%
\bibitem [{\citenamefont {Korshunov}(1986)}]{Korshunov1986}%
  \BibitemOpen
  \bibfield  {author} {\bibinfo {author} {\bibfnamefont {S.~E.}\ \bibnamefont {Korshunov}},\ }\bibfield  {title} {\bibinfo {title} {Phase diagram of the modified {$XY$} model},\ }\href {https://doi.org/10.1088/0022-3719/19/23/007} {\bibfield  {journal} {\bibinfo  {journal} {J. Phys. C: Solid State Phys.}\ }\textbf {\bibinfo {volume} {19}},\ \bibinfo {pages} {4427} (\bibinfo {year} {1986})}\BibitemShut {NoStop}%
\bibitem [{\citenamefont {Carpenter}\ and\ \citenamefont {Chalker}(1989)}]{CarpenterChalker1989}%
  \BibitemOpen
  \bibfield  {author} {\bibinfo {author} {\bibfnamefont {D.~B.}\ \bibnamefont {Carpenter}}\ and\ \bibinfo {author} {\bibfnamefont {J.~T.}\ \bibnamefont {Chalker}},\ }\bibfield  {title} {\bibinfo {title} {The phase diagram of a generalised {$XY$} model},\ }\href {https://doi.org/10.1088/0953-8984/1/30/004} {\bibfield  {journal} {\bibinfo  {journal} {J. Phys.: Condens. Matter}\ }\textbf {\bibinfo {volume} {1}},\ \bibinfo {pages} {4907} (\bibinfo {year} {1989})}\BibitemShut {NoStop}%
\bibitem [{\citenamefont {Poderoso}\ \emph {et~al.}(2011)\citenamefont {Poderoso}, \citenamefont {Arenzon},\ and\ \citenamefont {Levin}}]{Poderoso2011}%
  \BibitemOpen
  \bibfield  {author} {\bibinfo {author} {\bibfnamefont {F.~C.}\ \bibnamefont {Poderoso}}, \bibinfo {author} {\bibfnamefont {J.~J.}\ \bibnamefont {Arenzon}},\ and\ \bibinfo {author} {\bibfnamefont {Y.}~\bibnamefont {Levin}},\ }\bibfield  {title} {\bibinfo {title} {New ordered phases in a class of generalized {$XY$} models},\ }\href {https://doi.org/10.1103/PhysRevLett.106.067202} {\bibfield  {journal} {\bibinfo  {journal} {Phys. Rev. Lett.}\ }\textbf {\bibinfo {volume} {106}},\ \bibinfo {pages} {067202} (\bibinfo {year} {2011})}\BibitemShut {NoStop}%
\bibitem [{\citenamefont {Shi}\ \emph {et~al.}(2011)\citenamefont {Shi}, \citenamefont {Lamacraft},\ and\ \citenamefont {Fendley}}]{Shi2011}%
  \BibitemOpen
  \bibfield  {author} {\bibinfo {author} {\bibfnamefont {Y.}~\bibnamefont {Shi}}, \bibinfo {author} {\bibfnamefont {A.}~\bibnamefont {Lamacraft}},\ and\ \bibinfo {author} {\bibfnamefont {P.}~\bibnamefont {Fendley}},\ }\bibfield  {title} {\bibinfo {title} {Boson pairing and unusual criticality in a generalized {XY} model},\ }\href {https://doi.org/10.1103/PhysRevLett.107.240601} {\bibfield  {journal} {\bibinfo  {journal} {Phys. Rev. Lett.}\ }\textbf {\bibinfo {volume} {107}},\ \bibinfo {pages} {240601} (\bibinfo {year} {2011})}\BibitemShut {NoStop}%
\bibitem [{\citenamefont {H{\"u}bscher}\ and\ \citenamefont {Wessel}(2013)}]{HubscherWessel2013}%
  \BibitemOpen
  \bibfield  {author} {\bibinfo {author} {\bibfnamefont {D.~M.}\ \bibnamefont {H{\"u}bscher}}\ and\ \bibinfo {author} {\bibfnamefont {S.}~\bibnamefont {Wessel}},\ }\bibfield  {title} {\bibinfo {title} {Stiffness jump in the generalized {$XY$} model on the square lattice},\ }\href {https://doi.org/10.1103/PhysRevE.87.062112} {\bibfield  {journal} {\bibinfo  {journal} {Phys. Rev. E}\ }\textbf {\bibinfo {volume} {87}},\ \bibinfo {pages} {062112} (\bibinfo {year} {2013})}\BibitemShut {NoStop}%
\bibitem [{\citenamefont {Serna}\ \emph {et~al.}(2017)\citenamefont {Serna}, \citenamefont {Chalker},\ and\ \citenamefont {Fendley}}]{Serna2017}%
  \BibitemOpen
  \bibfield  {author} {\bibinfo {author} {\bibfnamefont {P.}~\bibnamefont {Serna}}, \bibinfo {author} {\bibfnamefont {J.~T.}\ \bibnamefont {Chalker}},\ and\ \bibinfo {author} {\bibfnamefont {P.}~\bibnamefont {Fendley}},\ }\bibfield  {title} {\bibinfo {title} {Deconfinement transitions in a generalised {XY} model},\ }\href {https://doi.org/10.1088/1751-8121/aa89a1} {\bibfield  {journal} {\bibinfo  {journal} {J. Phys. A: Math. Theor.}\ }\textbf {\bibinfo {volume} {50}},\ \bibinfo {pages} {424003} (\bibinfo {year} {2017})}\BibitemShut {NoStop}%
\bibitem [{\citenamefont {Fertig}\ \emph {et~al.}(1977)\citenamefont {Fertig}, \citenamefont {Johnston}, \citenamefont {DeLong}, \citenamefont {McCallum}, \citenamefont {Maple},\ and\ \citenamefont {Matthias}}]{Fertig1977}%
  \BibitemOpen
  \bibfield  {author} {\bibinfo {author} {\bibfnamefont {W.~A.}\ \bibnamefont {Fertig}}, \bibinfo {author} {\bibfnamefont {D.~C.}\ \bibnamefont {Johnston}}, \bibinfo {author} {\bibfnamefont {L.~E.}\ \bibnamefont {DeLong}}, \bibinfo {author} {\bibfnamefont {R.~W.}\ \bibnamefont {McCallum}}, \bibinfo {author} {\bibfnamefont {M.~B.}\ \bibnamefont {Maple}},\ and\ \bibinfo {author} {\bibfnamefont {B.~T.}\ \bibnamefont {Matthias}},\ }\bibfield  {title} {\bibinfo {title} {Destruction of superconductivity at the onset of long-range magnetic order in the compound {$\mathrm{ErRh}_4\mathrm{B}_4$}},\ }\href {https://doi.org/10.1103/PhysRevLett.38.987} {\bibfield  {journal} {\bibinfo  {journal} {Phys. Rev. Lett.}\ }\textbf {\bibinfo {volume} {38}},\ \bibinfo {pages} {987} (\bibinfo {year} {1977})}\BibitemShut {NoStop}%
\bibitem [{\citenamefont {Wu}\ \emph {et~al.}(2012)\citenamefont {Wu}, \citenamefont {Valls},\ and\ \citenamefont {Halterman}}]{WuValls2012}%
  \BibitemOpen
  \bibfield  {author} {\bibinfo {author} {\bibfnamefont {C.-T.}\ \bibnamefont {Wu}}, \bibinfo {author} {\bibfnamefont {O.~T.}\ \bibnamefont {Valls}},\ and\ \bibinfo {author} {\bibfnamefont {K.}~\bibnamefont {Halterman}},\ }\bibfield  {title} {\bibinfo {title} {Reentrant superconducting phase in conical-ferromagnet--superconductor nanostructures},\ }\href {https://doi.org/10.1103/PhysRevLett.108.117005} {\bibfield  {journal} {\bibinfo  {journal} {Phys. Rev. Lett.}\ }\textbf {\bibinfo {volume} {108}},\ \bibinfo {pages} {117005} (\bibinfo {year} {2012})}\BibitemShut {NoStop}%
\bibitem [{\citenamefont {Schupper}\ and\ \citenamefont {Shnerb}(2004)}]{Schupper2004}%
  \BibitemOpen
  \bibfield  {author} {\bibinfo {author} {\bibfnamefont {N.}~\bibnamefont {Schupper}}\ and\ \bibinfo {author} {\bibfnamefont {N.~M.}\ \bibnamefont {Shnerb}},\ }\bibfield  {title} {\bibinfo {title} {Spin model for inverse melting and inverse glass transition},\ }\href {https://doi.org/10.1103/PhysRevLett.93.037202} {\bibfield  {journal} {\bibinfo  {journal} {Phys. Rev. Lett.}\ }\textbf {\bibinfo {volume} {93}},\ \bibinfo {pages} {037202} (\bibinfo {year} {2004})}\BibitemShut {NoStop}%
\bibitem [{\citenamefont {Fisher}(1959)}]{Fisher1959}%
  \BibitemOpen
  \bibfield  {author} {\bibinfo {author} {\bibfnamefont {M.~E.}\ \bibnamefont {Fisher}},\ }\bibfield  {title} {\bibinfo {title} {Transformations of {Ising} models},\ }\href {https://doi.org/10.1103/PhysRev.113.969} {\bibfield  {journal} {\bibinfo  {journal} {Phys. Rev.}\ }\textbf {\bibinfo {volume} {113}},\ \bibinfo {pages} {969} (\bibinfo {year} {1959})}\BibitemShut {NoStop}%
\bibitem [{\citenamefont {Syozi}\ and\ \citenamefont {Nakano}(1968)}]{SyoziNakano1968}%
  \BibitemOpen
  \bibfield  {author} {\bibinfo {author} {\bibfnamefont {I.}~\bibnamefont {Syozi}}\ and\ \bibinfo {author} {\bibfnamefont {H.}~\bibnamefont {Nakano}},\ }\bibfield  {title} {\bibinfo {title} {Appearance of two ferromagnetic phases separated by a paramagnetic phase in certain kinds of decorated {Ising} lattice},\ }\href {https://doi.org/10.1143/PTP.40.237} {\bibfield  {journal} {\bibinfo  {journal} {Prog. Theor. Phys.}\ }\textbf {\bibinfo {volume} {40}},\ \bibinfo {pages} {237} (\bibinfo {year} {1968})}\BibitemShut {NoStop}%
\bibitem [{\citenamefont {Ma{\v t}a{\v s}ovsk{\'a}}\ and\ \citenamefont {Ja{\v s}{\v c}ur}(2007)}]{Matasovska2007}%
  \BibitemOpen
  \bibfield  {author} {\bibinfo {author} {\bibfnamefont {S.}~\bibnamefont {Ma{\v t}a{\v s}ovsk{\'a}}}\ and\ \bibinfo {author} {\bibfnamefont {M.}~\bibnamefont {Ja{\v s}{\v c}ur}},\ }\bibfield  {title} {\bibinfo {title} {Reentrant transitions and multicompensation phenomena of the exactly solvable mixed spin-{$S$} and spin-{$1/2$} {Ising} model on decorated planar lattices},\ }\href {https://doi.org/10.1016/j.physa.2007.05.004} {\bibfield  {journal} {\bibinfo  {journal} {Physica A}\ }\textbf {\bibinfo {volume} {383}},\ \bibinfo {pages} {339} (\bibinfo {year} {2007})}\BibitemShut {NoStop}%
\bibitem [{\citenamefont {Mermin}\ and\ \citenamefont {Wagner}(1966)}]{Mermin1966}%
  \BibitemOpen
  \bibfield  {author} {\bibinfo {author} {\bibfnamefont {N.~D.}\ \bibnamefont {Mermin}}\ and\ \bibinfo {author} {\bibfnamefont {H.}~\bibnamefont {Wagner}},\ }\bibfield  {title} {\bibinfo {title} {Absence of ferromagnetism or antiferromagnetism in one- or two-dimensional isotropic {Heisenberg} models},\ }\href {https://doi.org/10.1103/PhysRevLett.17.1133} {\bibfield  {journal} {\bibinfo  {journal} {Phys. Rev. Lett.}\ }\textbf {\bibinfo {volume} {17}},\ \bibinfo {pages} {1133} (\bibinfo {year} {1966})}\BibitemShut {NoStop}%
\bibitem [{Sup()}]{SupplementalMaterial}%
  \BibitemOpen
  \href@noop {} {}\bibinfo {note} {See Supplemental Material at [URL will be inserted by publisher] for the exact bond trace and branch mappings, Monte Carlo methods, phase-boundary and finite-size analyses, stiffness measurements, and the two defect free energies. It includes additional references~\cite{Swendsen1987,Weber1988,Hsieh2013,WuGuo2012,Onsager1944}}\BibitemShut {NoStop}%
\bibitem [{\citenamefont {Houtappel}(1950)}]{Houtappel1950}%
  \BibitemOpen
  \bibfield  {author} {\bibinfo {author} {\bibfnamefont {R.~M.~F.}\ \bibnamefont {Houtappel}},\ }\bibfield  {title} {\bibinfo {title} {Order-disorder in hexagonal lattices},\ }\href {https://doi.org/10.1016/0031-8914(50)90130-3} {\bibfield  {journal} {\bibinfo  {journal} {Physica}\ }\textbf {\bibinfo {volume} {16}},\ \bibinfo {pages} {425} (\bibinfo {year} {1950})}\BibitemShut {NoStop}%
\bibitem [{\citenamefont {Liaw}\ \emph {et~al.}(1999)\citenamefont {Liaw}, \citenamefont {Huang}, \citenamefont {Lin},\ and\ \citenamefont {Wu}}]{Liaw1999}%
  \BibitemOpen
  \bibfield  {author} {\bibinfo {author} {\bibfnamefont {T.-M.}\ \bibnamefont {Liaw}}, \bibinfo {author} {\bibfnamefont {M.-C.}\ \bibnamefont {Huang}}, \bibinfo {author} {\bibfnamefont {S.~C.}\ \bibnamefont {Lin}},\ and\ \bibinfo {author} {\bibfnamefont {M.-C.}\ \bibnamefont {Wu}},\ }\bibfield  {title} {\bibinfo {title} {Scaling functions of interfacial tensions for a class of {Ising} cylinders},\ }\href {https://doi.org/10.1103/PhysRevB.60.12994} {\bibfield  {journal} {\bibinfo  {journal} {Phys. Rev. B}\ }\textbf {\bibinfo {volume} {60}},\ \bibinfo {pages} {12994} (\bibinfo {year} {1999})}\BibitemShut {NoStop}%
\bibitem [{\citenamefont {Wu}(2006)}]{Wu2006}%
  \BibitemOpen
  \bibfield  {author} {\bibinfo {author} {\bibfnamefont {M.-C.}\ \bibnamefont {Wu}},\ }\bibfield  {title} {\bibinfo {title} {Exact finite-size scaling functions for the interfacial tensions of the {Ising} model on planar lattices},\ }\href {https://doi.org/10.1103/PhysRevE.73.046135} {\bibfield  {journal} {\bibinfo  {journal} {Phys. Rev. E}\ }\textbf {\bibinfo {volume} {73}},\ \bibinfo {pages} {046135} (\bibinfo {year} {2006})}\BibitemShut {NoStop}%
\bibitem [{\citenamefont {Nelson}\ and\ \citenamefont {Kosterlitz}(1977)}]{Nelson1977}%
  \BibitemOpen
  \bibfield  {author} {\bibinfo {author} {\bibfnamefont {D.~R.}\ \bibnamefont {Nelson}}\ and\ \bibinfo {author} {\bibfnamefont {J.~M.}\ \bibnamefont {Kosterlitz}},\ }\bibfield  {title} {\bibinfo {title} {Universal jump in the superfluid density of two-dimensional superfluids},\ }\href {https://doi.org/10.1103/PhysRevLett.39.1201} {\bibfield  {journal} {\bibinfo  {journal} {Phys. Rev. Lett.}\ }\textbf {\bibinfo {volume} {39}},\ \bibinfo {pages} {1201} (\bibinfo {year} {1977})}\BibitemShut {NoStop}%
\bibitem [{\citenamefont {Weber}\ and\ \citenamefont {Minnhagen}(1988)}]{Weber1988}%
  \BibitemOpen
  \bibfield  {author} {\bibinfo {author} {\bibfnamefont {H.}~\bibnamefont {Weber}}\ and\ \bibinfo {author} {\bibfnamefont {P.}~\bibnamefont {Minnhagen}},\ }\bibfield  {title} {\bibinfo {title} {{Monte Carlo} determination of the critical temperature for the two-dimensional {XY} model},\ }\href {https://doi.org/10.1103/PhysRevB.37.5986} {\bibfield  {journal} {\bibinfo  {journal} {Phys. Rev. B}\ }\textbf {\bibinfo {volume} {37}},\ \bibinfo {pages} {5986} (\bibinfo {year} {1988})}\BibitemShut {NoStop}%
\bibitem [{\citenamefont {Hsieh}\ \emph {et~al.}(2013)\citenamefont {Hsieh}, \citenamefont {Kao},\ and\ \citenamefont {Sandvik}}]{Hsieh2013}%
  \BibitemOpen
  \bibfield  {author} {\bibinfo {author} {\bibfnamefont {Y.-D.}\ \bibnamefont {Hsieh}}, \bibinfo {author} {\bibfnamefont {Y.-J.}\ \bibnamefont {Kao}},\ and\ \bibinfo {author} {\bibfnamefont {A.~W.}\ \bibnamefont {Sandvik}},\ }\bibfield  {title} {\bibinfo {title} {Finite-size scaling method for the {Berezinskii--Kosterlitz--Thouless} transition},\ }\href {https://doi.org/10.1088/1742-5468/2013/09/P09001} {\bibfield  {journal} {\bibinfo  {journal} {J. Stat. Mech.}\ ,\ \bibinfo {pages} {P09001}} (\bibinfo {year} {2013})}\BibitemShut {NoStop}%
\bibitem [{\citenamefont {Swendsen}\ and\ \citenamefont {Wang}(1987)}]{Swendsen1987}%
  \BibitemOpen
  \bibfield  {author} {\bibinfo {author} {\bibfnamefont {R.~H.}\ \bibnamefont {Swendsen}}\ and\ \bibinfo {author} {\bibfnamefont {J.-S.}\ \bibnamefont {Wang}},\ }\bibfield  {title} {\bibinfo {title} {Nonuniversal critical dynamics in {Monte Carlo} simulations},\ }\href {https://doi.org/10.1103/PhysRevLett.58.86} {\bibfield  {journal} {\bibinfo  {journal} {Phys. Rev. Lett.}\ }\textbf {\bibinfo {volume} {58}},\ \bibinfo {pages} {86} (\bibinfo {year} {1987})}\BibitemShut {NoStop}%
\bibitem [{\citenamefont {Onsager}(1944)}]{Onsager1944}%
  \BibitemOpen
  \bibfield  {author} {\bibinfo {author} {\bibfnamefont {L.}~\bibnamefont {Onsager}},\ }\bibfield  {title} {\bibinfo {title} {Crystal statistics. {I}. {A} two-dimensional model with an order-disorder transition},\ }\href {https://doi.org/10.1103/PhysRev.65.117} {\bibfield  {journal} {\bibinfo  {journal} {Phys. Rev.}\ }\textbf {\bibinfo {volume} {65}},\ \bibinfo {pages} {117} (\bibinfo {year} {1944})}\BibitemShut {NoStop}%
\bibitem [{\citenamefont {Wu}\ and\ \citenamefont {Guo}(2012)}]{WuGuo2012}%
  \BibitemOpen
  \bibfield  {author} {\bibinfo {author} {\bibfnamefont {F.~Y.}\ \bibnamefont {Wu}}\ and\ \bibinfo {author} {\bibfnamefont {W.}~\bibnamefont {Guo}},\ }\bibfield  {title} {\bibinfo {title} {Critical manifold of the {Potts} model: Exact results and homogeneity approximation},\ }\href {https://doi.org/10.1103/PhysRevE.86.020101} {\bibfield  {journal} {\bibinfo  {journal} {Phys. Rev. E}\ }\textbf {\bibinfo {volume} {86}},\ \bibinfo {pages} {020101(R)} (\bibinfo {year} {2012})}\BibitemShut {NoStop}%
\end{thebibliography}%

\end{document}


\makeatletter\immediate\write\@auxout{\string\citation{apsrev42Control}}\makeatother
\nocite{Shi2011,Serna2017,Houtappel1950,Liaw1999,Wu2006,Nelson1977,Weber1988,Hsieh2013,Swendsen1987,Onsager1944,WuGuo2012}
\maketitle
\tableofcontents

\section{Model and bond-state summation}\label{sec:geometry}\label{sec:trace}
The sites of the periodic triangular lattice lie at \(\bm r=x\bm a_1+y\bm a_2\),
with \(\bm a_1=(1,0)\), \(\bm a_2=(1/2,\sqrt3/2)\), and \(x,y=0,\ldots,L-1\).
Counting one bond per site along each of \(\bm a_1,\bm a_2,\bm a_2-\bm a_1\)
gives \(N_b=3N_s\) bonds for \(N_s=L^2\) sites. The area of the periodic
rhombus is \(A=\sqrt3 L^2/2\). We set the lattice spacing and \(k_B\) to one
and measure energies in units of \(J_0+\lambda\).

Each site carries a phase \(\theta_i\in[0,2\pi)\), and each nearest-neighbor
bond \(b=\langle ij\rangle\) carries an annealed bond variable \(n_b=0,1\).
For a phase difference \(\delta_b=\theta_i-\theta_j\), the energy required
to excite the bond is \(a_b=\varepsilon+\lambda(1-\cos\delta_b)\)
[main-text Eq.~\textup{(2)}]. At fixed phases the bonds fluctuate
independently, with excitation probability
\(p_b=\Pr(n_b=1\mid\theta)=(1+e^{a_b/T})^{-1}\).
The sum over bond states gives the angular partition function
\begin{align}
 Z&=\int\dd\mu(\theta)\sum_{\{n_b\}}e^{-H(\theta,n)/T}
   =\int\dd\mu(\theta)\prod_b w_b(\delta_b),\qquad
 \dd\mu(\theta)=\prod_i\frac{\dd\theta_i}{2\pi},\nonumber\\
 w_b(\delta_b)&=\exp[-f_b(\delta_b,T)/T],\qquad
 f_b=-J_0\cos\delta_b-J_2\cos2\delta_b-T\ln(1+e^{-a_b/T}).
 \label{eq:feff}
\end{align}
The traced weight determines the angular distribution. Energy measurements
require the conditional mean of the microscopic Hamiltonian,
\begin{equation}
 E_c(\theta)=\frac{\partial}{\partial\beta}
       \left[\beta\sum_b f_b(\delta_b,T)\right]
   =\sum_b[-J_0\cos\delta_b-J_2\cos2\delta_b+a_b p_b],
 \qquad \beta=1/T .\label{eq:Ebar}
\end{equation}
Thus \(\langle H\rangle=\langle E_c\rangle\). The conditional bond
variance \(p_b(1-p_b)\) enters the helicity modulus in Sec.~\ref{sec:bkt}.

\section{Ising mapping and reentrant order}\label{sec:exact}
At fixed finite \(T\), taking \(J_2\to\infty\) locks the charge-\(4e\) phase
\(\phi_i=2\theta_i\bmod 2\pi\) across the lattice. The remaining fields satisfy
\(e^{i\theta_i}=\sigma_i e^{i\theta_0}\), where \(\sigma_i=\pm1\).
For \(q=\sigma_i\sigma_j\), the two nearest-neighbor weights are
\begin{equation}
 W_q=e^{(J_2+J_0 q)/T}
       [1+e^{-[\varepsilon+\lambda(1-q)]/T}]
       =\sqrt{W_+W_-}\,e^{Kq}.
\end{equation}
The common factor is independent of the Ising spins. Their nearest-neighbor
energy coupling \(J_{\rm I}(T)\) follows from the ratio of the two weights.
In units of temperature it is
\begin{equation}
 K(T)=\frac{J_{\rm I}(T)}{T}
 =\frac{J_0}{T}+\frac12\ln\frac{1+e^{-\varepsilon/T}}
                                  {1+e^{-(\varepsilon+2\lambda)/T}}.
 \label{eq:Kexact}
\end{equation}
For \(J_0,\varepsilon>0\), \(K(T)\sim J_0/T\) as \(T\to0\). At high
temperature \(J_{\rm I}\to J_0+\lambda/2\), so \(K\) tends to zero.
The critical coupling of the triangular Ising ferromagnet is
\(K_c=\tfrac14\ln3\)~\cite{Houtappel1950}. For \(J_0=0.025\),
\(\lambda=0.975\), and \(\varepsilon=0.235\),
the three solutions of \(K(T)=K_c\) lie at
\(T_{1,2,3}^{(\infty)}\simeq 0.120,\ 0.330,\ 0.852\), respectively.
The slopes \(K'(T)\) there are \(-0.739,\ +0.112,\ -0.0721\).
All three crossings thus survive sufficiently small changes in the couplings.

The activation cost on the critical curve can be obtained directly from the
ratio of bond weights. With \(u=e^{2K_c-2J_0/T}\) and \(z=e^{-2\lambda/T}\),
\begin{equation}
 \varepsilon_c(T)=T\ln\frac{1-uz}{u-1},\qquad
 1<u<\frac{2}{1+z}.\label{eq:criticalcurve}
\end{equation}
These inequalities select finite positive \(\varepsilon_c\).
Here \(p_+\) and \(p_-\) are the conditional excitation probabilities at
\(\delta_b=0\) and \(\pi\), respectively [main-text Eq.~\textup{(3)}].
Since
\(\partial_\varepsilon K=-(p_+-p_-)/(2T)<0\), the ordered side lies at
\(\varepsilon<\varepsilon_c(T)\) wherever Eq.~\eqref{eq:criticalcurve} applies.
Two crossings merge where \(K=K_c\) and \(\partial_T K=0\).
A simple fold also requires \(\partial_T^2K\ne0\).

The entropy difference between aligned and antialigned bonds limits the increase in \(K\) on heating. For an excitation
of cost \(a\) and equilibrium probability \(p\), differentiating \(T\ln(1+e^{-a/T})\) with respect to \(T\) gives the two-state entropy
\(s(p)=-p\ln p-(1-p)\ln(1-p)\). Consequently
\begin{equation}
 2\,\frac{\dd J_{\rm I}}{\dd T}=2[K+TK']=s(p_+)-s(p_-).
\end{equation}
For finite \(T\), \(0<p_-<p_+<1/2\), so this difference lies strictly between
zero and \(\ln2\). The energy coupling grows monotonically. To restore Ising
branch order upon heating, \(K\) must cross \(K_c\) from below. There
\(K'\ge0\), with strict inequality for a transverse crossing. Consequently
\(2K_c\le s(p_+)-s(p_-)<\ln2\). The entropy bound \(2K_c<\ln2\) is
therefore necessary for recovery. Whether recovery occurs for a given set of
couplings depends on the crossings of Eq.~\eqref{eq:Kexact}.
At the middle triangular-lattice crossing the entropy difference is about
0.62, exceeding \(2K_c\simeq0.55\). For the square-lattice nearest-neighbor
Ising ferromagnet, \(K_c=\tfrac12\ln(1+\sqrt2)\)~\cite{Onsager1944},
which exceeds the bound for this constrained two-state model.

A bond cut by an Ising domain wall has a bare free-energy cost
\(-T\ln(W_-/W_+)=2J_{\rm I}\) [main-text Eq.~\textup{(11)}].
The thermodynamic interfacial tension \(\tau_{\rm I}\) also accounts for the
configurational entropy of the wall. It vanishes for \(K\le K_c\) and is
positive for \(K>K_c\). For a fixed orientation in two dimensions,
\(\tau_{\rm I}/T\sim\xi^{-1}\propto K-K_c\) on approaching the transition
from the ordered side, where \(\xi\) is the Ising correlation length~\cite{Liaw1999,Wu2006}.
Near each transverse crossing
this gives \(\tau_{\rm I}/T\propto |K'(T_i^{(\infty)})|\,|T-T_i^{(\infty)}|\).
The charge-\(4e\) field is rigid throughout this limit. Its eventual melting
and the separation free energy of half vortices require the fluctuating
continuous-angle model at finite \(J_2\).

\section{Monte Carlo methods and statistical analysis}\label{sec:MC}\label{sec:diagnostics}

\paragraph{Equilibrium updates.}
We simulate the periodic triangular lattice with annealed bond variables.
Each sweep starts with a heat-bath draw of every \(n_b\) from its conditional
probability \(p_b\) in Sec.~\ref{sec:geometry}.
For the resulting \(J_b=J_0+\lambda n_b\), rotating one endpoint of a bond by
\(\pi\) costs \(\Delta E_b=2J_b\cos(\theta_i-\theta_j)\). We connect the two
sites with probability \(1-\exp[-\max(0,\Delta E_b)/T]\). Each connected
cluster rotates by \(\pi\) with probability \(1/2\)~\cite{Swendsen1987}.
Two single-site Metropolis passes update the continuous angles. The first
proposes a uniform increment of half-width \(2\sqrt{T/(24J_2)}\). The second
reflects the phase about the local axis of its neighbors' charge-\(4e\)
field. This reflection preserves the second-harmonic energy.

In the initial critical scans and the \(T=0.19\) runs at \(L=128,256,512\),
we supplement these moves with a global twist of \(\pm\pi\) along a
random lattice coordinate every 16 sweeps. The Metropolis test uses the
bond-summed weight [Eq.~\eqref{eq:feff}], including bonds across the periodic
seam. In the later phase-diagram scans, additional temperature scans for
Fig.~3{}(c), and large-size recovery runs, we choose the twist
magnitude from \(\pi\) and \(2\pi\) with equal probability. These runs also
include a reflection-cluster move in each sweep. For a uniform axis angle
\(\alpha\in[0,\pi)\), put \(v_i=\sin(\theta_i-\alpha)\) and
\(w_i=\sin[2(\theta_i-\alpha)]\). The bond-energy change for reflecting one
endpoint is \(2(J_bv_iv_j+J_2w_iw_j)\). With this cost in the connection
probability above, each cluster is independently reflected about \(\alpha\)
with probability \(1/2\).

Every 64 sweeps the later runs also propose a half twist in 32 increments
about a randomly placed cut. We alternate the increments with reversible
\(\pi\)-cluster moves in the bond-summed ensemble, starting and ending with a
cluster move. The reverse proposal applies the opposite twist. The acceptance
probability is \(\min(1,e^{-W/T})\), where \(W\) sums the changes in the
bond-summed free energy at the twist increments. After these angular moves
the next sweep samples fresh bond states. We checked the update schemes by
small-lattice enumeration, comparison with single-site Metropolis sampling,
and detailed-balance tests of the reflection and twist proposals.

\paragraph{Equilibration and errors.}
Each parameter point uses independently seeded runs with four initial
conditions: aligned phases, random \(0/\pi\) phases, a winding profile
\(\pi x/L\) with random \(0/\pi\) shifts, and uniform random phases.
For the phase-interior measurements we record every eighth sweep, with
\(16384\) to \(131072\) sweeps per run. Critical scans typically have \(32768\)
or \(65536\) sweeps. We compare the different initial conditions and analyze
each history with two equilibration cutoffs. At low temperature an initial
winding or vortex can survive the earlier cutoff. Near charge-\(4e\)
melting, vortices also appear in equilibrium. We use their persistence,
dependence on initial conditions, and the drift of the observables to
identify histories that have yet to equilibrate. The projected defect
ensembles in Sec.~\ref{sec:defects} impose separate constraints.

For each monitored observable we estimate the integrated autocorrelation time
\(\tau_{\rm int}\) by pairing the autocorrelations at lags \(1,2\), then \(3,4\), and so on.
We retain the initial sequence of positive pair sums and replace each by its running minimum.
In units of the interval between saved measurements, the lower bound on \(\tau_{\rm int}\) is \(1/2\).
For \(n\) measurements the effective
sample size (ESS) is \(n/(2\tau_{\rm int})\). We flag a point if a run has
ESS below 100, if the split-chain statistic \(\widehat R\) exceeds 1.05, or
if the means of the first and last halves of a retained run differ by more
than three autocorrelation-adjusted standard errors. Monitored quantities
include energy, bond occupation, order-parameter moments, correlations,
currents, winding, and vortex density. For derived observables we resample
contiguous time blocks separately within each run, drawing all observables
from a given block together. The two cutoffs sample overlapping portions of
a history. The quoted bootstrap intervals are conditional on the chosen
cutoff, fit form, and size range.

\section{Correlations and phase boundaries}\label{sec:observables}

For \(m=1,2\), the single-configuration estimators and their equilibrium means are
\begin{align}
 \widehat G_m(r;\theta)&=\frac{1}{3N_s}\sum_{i,\mu}
 \cos[m(\theta_i-\theta_{i+r\bm d_\mu})],
 &G_{2me}(r)&=\langle\widehat G_m(r;\theta)\rangle,\\
 \widehat M_m(\theta)&=\left|\frac{1}{N_s}\sum_j e^{im\theta_j}\right|^2,
 &M_m&=\langle\widehat M_m\rangle,\qquad S_{2me}(L,T)=N_s M_m .
\end{align}
The sum over \(\mu\) averages the three nearest-neighbor directions. A correlation decaying as \(r^{-\eta_{2me}}\) gives \(S_{2me}\propto L^{2-\eta_{2me}}\). For short-range correlations \(S_{2me}\) saturates once \(L\) exceeds the correlation length. Both channels show algebraic correlations in SC, only the \(4e\) channel retains them in the charge-\(4e\) phase, and both become short ranged in the normal phase.

\begin{table}[htbp]\centering\small
\caption{Structure factors in Fig.~3{}(a). Each temperature has two columns, the charge-\(2e\) structure factor and the charge-\(4e\) structure factor per site. Parentheses give one block-bootstrap standard error in the last digits.}\label{tab:anchor_channels}
\setlength{\tabcolsep}{5pt}
\begin{tabular}{r rr rr rr}\toprule
&\multicolumn{2}{c}{\(T=0.19\)}&\multicolumn{2}{c}{\(T=0.56\)}&\multicolumn{2}{c}{\(T=1.2\)}\\
\cmidrule(lr){2-3}\cmidrule(lr){4-5}\cmidrule(lr){6-7}
\(L\)&\(S_{2e}\)&\(S_{4e}/L^2\)&\(S_{2e}\)&\(S_{4e}/L^2\)&\(S_{2e}\)&\(S_{4e}/L^2\)\\\midrule
32 &106.0(14)&0.977606(56)&623.2(16)&0.93530(13)&36.36(56)&0.86141(25)\\
64 &111.3(17)&0.973834(55)&2455.4(37)&0.92421(17)&34.82(61)&0.83832(27)\\
128&115.08(63)&0.969804(32)&9788.5(66)&0.91331(15)&36.43(53)&0.81623(27)\\
256&113.88(54)&0.966032(31)&39019(14)&0.90246(18)&35.74(51)&0.79449(29)\\
512&114.51(71)&0.962074(47)&155511(28)&0.89168(21)&35.68(52)&0.77383(32)\\\bottomrule
\end{tabular}
\end{table}

We test each structure-factor channel separately at both equilibration cutoffs. Each of the four chains must pass the ESS and drift criteria in Sec.~\ref{sec:MC}, and the chains must pass the split-\(\widehat R\) test. For each measured moment we require at least eight blocks per run and a minimum block duration of ten autocorrelation times. Estimates from the two cutoffs must agree within three times the sum of their standard errors. All 30 channel estimates pass at both cutoffs. The largest split-\(\widehat R\) is 1.004, and the per-chain ESS exceeds 1044 for \(2e\) and 348 for \(4e\). Each error in Table~\ref{tab:anchor_channels} uses 4096 bootstrap samples from 32 blocks per chain, weighted by their measurement counts.

We discard 4096 sweeps for \(L=32,64\) and 16384 for larger sizes, except at \(T=0.19,L=128\), where we discard 32768. At that point, \(G_{2e}(64)\) still shows drift in the run initialized with a winding profile. The charge-\(2e\) structure factor gives \(115.45(55)\) at the earlier cutoff and \(115.08(63)\) at the later cutoff.

Figure~3{}(b) shows signed distance-dependent estimates on the \(L=512\) lattice. At \(r=L/2\), \(G_{2e}\) is \((-1.85\pm2.33)\times10^{-5}\), \(0.59258(11)\), and \((-3.68\pm4.10)\times10^{-5}\) at \(T=0.19,0.56,1.2\), respectively. The errors refer to \(G_{2e}\) itself. Open symbols at the two outer temperatures indicate that the denominator of the real-space ratio \(R_1=G_{2e}(L/2)/G_{2e}(L/4)\) lies within three standard errors of zero. This ratio can therefore be unresolved even when both structure factors have well-determined errors.

In Fig.~3{}(c), size pairs from \(32/64\) to \(256/512\) share a grid of 45 temperatures from \(0.08\) to \(6\). Including additional measurements at matched temperatures gives 305 size pairs and 610 charge-channel ratios across 91 distinct temperatures. In each channel \(R=R_{2me}(L,T)=S_{2me}(2L,T)/S_{2me}(L,T)\), with \(L\) the smaller size. Independent runs at the two sizes give \((\delta R/R)^2=[\delta S(2L)/S(2L)]^2+[\delta S(L)/S(L)]^2\) to first order. Adjacent size pairs share one measurement. Open symbols identify 134 estimates with an equilibration flag, four with an effective-sample flag, and two with only one usable cut at one size. The latter two are the \(256/512\) ratios at \(T=0.23\), where an initial defect survives the first cut on the larger lattice. The defect clears before the later cutoff at 24576 sweeps (minimum ESS 122). Along this scan, a ratio near \(2^{2-\eta_{2me}}\) and a ratio near one distinguish algebraic and short-range correlations.

The charge-\(2e\) boundaries in main-text Fig.~1{}(b) are estimated from the size crossings described below. At \(\varepsilon=0.235\), the \(L=128,256\) crossings occur at \(T\simeq0.117,0.368,0.769\) for \(J_2=2\) and \(0.118,0.348,0.810\) for \(J_2=4\). These finite-size features support the same ordering sequence at both couplings. To draw the phase boundary we interpolate \(\varepsilon\) as a function of \(\ln T\) through the crossings, window closures, and outer temperature brackets using a shape-preserving cubic Hermite interpolant. The high-temperature boundary joins the stiffness estimates in Sec.~\ref{sec:bkt}.

\section{Finite-size scaling of charge-\texorpdfstring{\(2e\)}{2e} coherence}\label{sec:criticalfits}

\paragraph{Coherence boundaries.}
At a charge-\(2e\) coherence boundary, write \(e^{i\theta_i}=\sigma_i e^{i\phi_i/2}\) using a local choice of half-angle for \(\phi_i=2\theta_i\bmod 2\pi\). If critical Ising fluctuations factorize from the gapless Gaussian fluctuations of \(\phi_i\), the Ising spin contributes \(1/4\) to the correlation exponent. The Gaussian exponent of \(e^{ia\phi}\) is proportional to \(a^2\), so the half-angle contributes \(\eta_{4e}/4\)~\cite{Shi2011,Serna2017}. The resulting prediction \(\eta_{2e}=1/4+\eta_{4e}/4\) gives two scale-invariant combinations,
\begin{equation}
 X(L,T)=L^{1/4}\frac{M_1}{M_2^{1/4}},\qquad
 R_{\rm I}(L,T)=\frac{\mathcal C_{\rm I}(L/2)}{\mathcal C_{\rm I}(L/4)},
 \qquad \mathcal C_{\rm I}(r)=\frac{G_{2e}(r)}{[G_{4e}(r)]^{1/4}} .
 \label{eq:SMX}
\end{equation}
The factor \(L^{1/4}\) in \(X\) fixes the Ising spin exponent in advance. Its crossing tests the factorization hypothesis, while the slope with temperature tests the separate thermal exponent \(1/\nu\). The real-space ratio \(R_{\rm I}\) has no explicit power of \(L\); its crossing value is fitted. Both observables use the same histories. To locate a phase boundary at fixed \(\varepsilon\), we bracket a sign change in \(X(2L,T)-X(L,T)\) and interpolate linearly between adjacent temperatures. On the \(\varepsilon=0.235\) cut we also fit multiple sizes together.

\begin{figure}[htbp]\centering
\includegraphics[width=.94\linewidth]{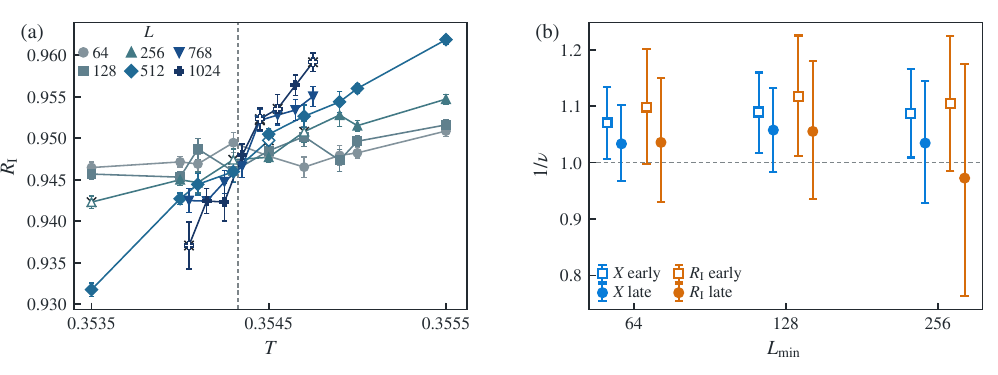}
\caption{Finite-size tests at the recovery boundary. (a) Real-space ratios \(R_{\rm I}\) at the later cutoff. The dashed line marks \(T_2\) from the fit to \(64\le L\le768\). Open, crossed points carry equilibration or effective-sample flags. (b) Thermal exponents and conditional 95\% bootstrap intervals as the minimum size is varied, including the accepted \(L=1024\) points. The fits use \(p=3\), \(\omega=2\), and adjacent blocks merged in pairs. Open squares and filled circles denote the earlier and later cutoffs of the same histories. The dashed line marks the Ising value \(1/\nu=1\).}\label{fig:SMscaling}
\end{figure}

We fit either observable with a polynomial in the thermal scaling variable, allowing a leading size correction,
\begin{equation}
 \mathcal O(L,T)=\sum_{k=0}^{p}a_k x^k+b\ell^{-\omega},
 \qquad x=\frac{T-T_c}{w}\ell^{1/\nu},\qquad \ell=L/128.
 \label{eq:thermalfit}
\end{equation}
The recovery fit uses \(0.3535\leq T\leq0.3555\), \(w=0.001\), \(p=3\), and fixed \(\omega=2\). Of the 45 late-cutoff measurements in Fig.~3{}(d), the counts passing the equilibration and effective-sample criteria are \(9/9,9/9,6/9,9/10,7/8\) for \(L=64,128,256,512,768\), respectively. The first two rows of Table~\ref{tab:recovery} give the fits to these 40 points. Of eight additional \(L=1024\) measurements, four pass, two have equilibration flags, and two have effective-sample flags. With only half of the largest-size points passing, we use the denser coverage through $L=768$ for the main fit. The last two rows show the sensitivity to including the four passing $L=1024$ points. At both maximum sizes, the late-cutoff thermal exponent is compatible with \(1/\nu=1\). At \(L_{\max}=1024\), changing to the earlier cut gives \(1/\nu=1.070\ [1.006,1.133]\) for \(X\). Figure~\ref{fig:SMscaling} shows the real-space crossing and the variation with the minimum size.

\begin{table}[htbp]\centering\small
\caption{Recovery-boundary fits using late-cutoff data, \(L_{\min}=64\) and Eq.~\eqref{eq:thermalfit}. Brackets span the 2.5th to 97.5th percentiles of 2048 chain-stratified bootstrap draws with adjacent blocks merged in pairs.}\label{tab:recovery}
\setlength{\tabcolsep}{4pt}
\begin{tabular}{rlrrr}\toprule
\(L_{\max}\)&Observable&\(T_2\)&\(1/\nu\)&\(\chi^2/\mathrm{dof}\)\\\midrule
768 &\(X\)&\(0.3543445\ [0.3543065,0.3543815]\)&\(1.030\ [0.959,1.105]\)&\(33.12/33\)\\
768 &\(R_{\rm I}\)&\(0.3543264\ [0.3542649,0.3543890]\)&\(1.026\ [0.912,1.156]\)&\(34.59/33\)\\
1024&\(X\)&\(0.354342\ [0.354313,0.354372]\)&\(1.033\ [0.967,1.103]\)&\(34.68/37\)\\
1024&\(R_{\rm I}\)&\(0.354327\ [0.354279,0.354375]\)&\(1.036\ [0.931,1.149]\)&\(36.60/37\)\\\bottomrule
\end{tabular}
\end{table}

For the first and third boundaries we use \(64\leq L\leq512\), a quadratic thermal polynomial, and fixed \(\omega=2\). The respective temperature windows are \([0.1178,0.1184]\) and \([0.794,0.802]\), with \(w=0.0003\) and \(0.004\). Representative late-cutoff \(X\) fits give \(T_1=0.118048\), \(1/\nu=0.975\ [0.960,0.991]\), and \(T_3=0.798027\), \(1/\nu=0.994\ [0.970,1.020]\). These brackets span the 16th to 84th percentiles of 240 bootstrap draws. At the third boundary a cubic fit of the real-space ratio gives \(1/\nu=0.937\) for \(L_{\min}=64\) and \(1.007\) for \(L_{\min}=128\). This shift illustrates the remaining size dependence in the thermal-exponent estimates.

\paragraph{Closure of the reentrant windows.}\label{sec:endpoints}
Near a closure, two zeros of the thermal scaling field \(g(e,t)\) merge. We use
\(e=(\varepsilon-\varepsilon_0)/\delta_\varepsilon\) and \(t=(T-T_0)/\delta_T\).
The four constants \((\varepsilon_0,\delta_\varepsilon,T_0,\delta_T)\) are
\((0.2067,0.0002,0.182,0.01)\) for the lower closure and
\((0.2643,0.0005,0.590,0.03)\) for the upper one.
We expand \(g\) in \(1,e,t,t^2,et,e^2,t^3\) and use \(1/\nu=1,\ \omega=2\).
The fit and the two conditions locating a closure are
\begin{align}
 X&=c_0+c_L\ell^{-2}+z+qz^2,\qquad z=\ell\,g(e,t),\label{eq:foldfits}\\
 g(e_*,t_*)&=0,\qquad \partial_t g(e_*,t_*)=0.\label{eq:fold}
\end{align}
A regular closure has \(\partial_e g\ne0\) and \(\partial_t^2g\ne0\). At the lower closure, including \(qz^2\) improves \(\chi^2/\mathrm{dof}\) from \(104.25/68\) to \(89.82/67\) for \(L_{\min}=128\). At the upper closure we retain the linear fit (\(q=0\)). Removing the smaller sizes tests the stability of both locations (Table~\ref{tab:folds}). At \(\varepsilon=0.26425,\ T=0.58\), a vortex in an \(L=1024\) hot-start history survives through sweep 52304. We exclude this affected measurement from the fits. Including only its tail after sweep 55296 shifts the upper-closure estimates by at most \(2.2\times10^{-5}\) in \(\varepsilon\) and \(2.1\times10^{-4}\) in \(T\) over the fits examined.

\begin{table}[htbp]\centering\small
\caption{Closures of the two reentrant temperature intervals in Fig.~1{}(b). Brackets give conditional 95\% block-bootstrap intervals at the late cutoff, with adjacent blocks merged in pairs and \(L_{\max}=1024\).}\label{tab:folds}
\begin{tabular}{lrrr}\toprule
Closure&\(L_{\min}\)&\(\varepsilon_*\)&\(T_*\)\\\midrule
Lower &128&\(0.206723\ [0.206717,0.206729]\)&\(0.183049\ [0.182872,0.183217]\)\\
Lower &512&\(0.206702\ [0.206665,0.206738]\)&\(0.183029\ [0.182818,0.183227]\)\\
Upper &128&\(0.264356\ [0.264342,0.264369]\)&\(0.592832\ [0.592102,0.593570]\)\\
Upper &512&\(0.264373\ [0.264290,0.264456]\)&\(0.592530\ [0.591732,0.593312]\)\\\bottomrule
\end{tabular}
\end{table}

\section{Charge-\texorpdfstring{$4e$}{4e} melting}\label{sec:bkt}

The helicity modulus measures the free-energy cost of a uniform gradient $\bm g$ of the charge-$2e$ phase. We replace $\delta_b=\theta_i-\theta_j$ by $\delta_b+\bm g\cdot\bm d_b$, where $\bm d_b$ is the physical unit vector along the oriented bond. After the bond trace in Eq.~\eqref{eq:feff}, the derivatives of its free energy are
\begin{align}
 f'_b&=(J_0+\lambda p_b)\sin\delta_b+2J_2\sin2\delta_b, \nonumber\\
 f''_b&=(J_0+\lambda p_b)\cos\delta_b+4J_2\cos2\delta_b
          -\lambda^2 p_b(1-p_b)\sin^2\delta_b/T .
\end{align}
The last term is the conditional current variance of the annealed bond. With $I_\mu=\sum_b f'_bd_{b\mu}$, the Cartesian helicity tensor and its isotropic part are
\begin{equation}
\rho_{\mu\nu}=\frac{1}{A}\left[
 \left\langle\sum_b f''_b d_{b\mu}d_{b\nu}\right\rangle
 -\frac{\langle I_\mu I_\nu\rangle-\langle I_\mu\rangle\langle I_\nu\rangle}{T}\right],
 \qquad \rho=\tfrac12(\rho_{xx}+\rho_{yy}),
 \label{eq:rho}
\end{equation}
with physical area $A=\sqrt{3}L^2/2$. Around a half vortex the smooth part of $\theta$ winds by $\pi$, so the Berezinskii--Kosterlitz--Thouless (BKT) jump of the bulk stiffness is $\rho/T=8/\pi$~\cite{Nelson1977}. The corresponding Gaussian coupling for $\phi=2\theta$ is $K_{4e}=\rho_{\rm bulk}/(4T)$.

To compare the measured $\rho_L$ with the bulk jump, we include windings of the charge-$4e$ field around the periodic cell. Let $D=(\bm a_1,\bm a_2)$, $a_0=|\det D|=\sqrt3/2$, and $\bm u_{\bm m}=D^{-T}\bm m$ for $\bm m\in\mathbb Z^2$. For a trial Gaussian coupling $\kappa$, the winding weights are $w_{\bm m}(\kappa)=\exp[-2\pi^2a_0\kappa|\bm u_{\bm m}|^2]$. Their response to a twist gives
\begin{equation}
 \frac{\rho_L}{T}=4K_{4e,L} f(K_{4e,L}),\qquad
 f(\kappa)=1-2\pi^2a_0\kappa\frac{\sum_{\bm m}|\bm u_{\bm m}|^2w_{\bm m}}
                              {\sum_{\bm m}w_{\bm m}} .
\end{equation}
Here $K_{4e,L}$ is the finite-size Gaussian coupling of the charge-$4e$ field. For this rhombus $f(2/\pi)=0.999957$. At the putative transition we compare the leading Weber--Minnhagen (WM) form~\cite{Weber1988} with a subleading logarithmic correction~\cite{Hsieh2013},
\begin{equation}
 K_{4e,L}^{\rm WM}=\frac{2}{\pi}\left(1+\frac{1}{2z}\right),\qquad
 K_{4e,L}^{\rm sub}=\frac{2}{\pi}\left(1+\frac{1}{2z+\ln z}\right),\qquad
 z=\ln L+C/2 ,
\end{equation}
where the constant $C$ is fitted separately for each form.

\begin{figure}[tbp]
 \centering
 \includegraphics[width=.94\linewidth]{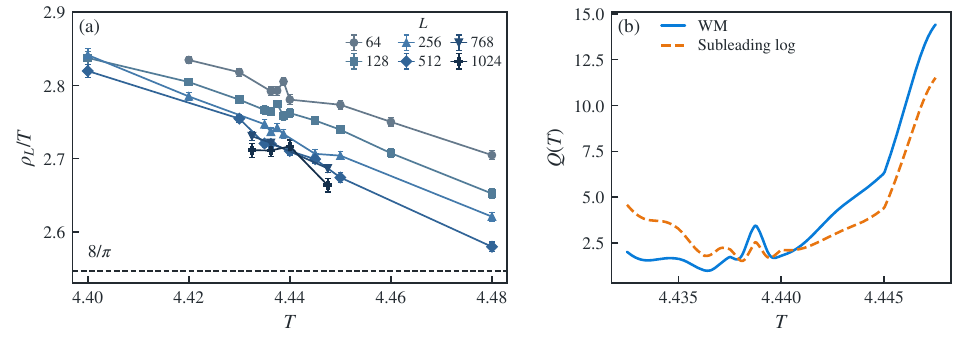}
 \caption{Charge-$4e$ melting at $J_2=3$ and $\varepsilon=0.235$.
 (a) Helicity modulus $\rho_L$ of the charge-$2e$ phase divided by temperature. Symbols show the measurements retained after the later cutoff, with one-standard-error bars. The horizontal line marks the bulk jump $8/\pi$ for half vortices. (b) Residual profiles $Q(T)$ for the two logarithmic forms. Each fit uses all six sizes and varies $C$ at each trial temperature.}
 \label{fig:stiffness}
\end{figure}

Figure~\ref{fig:stiffness}(a) shows the 45 of 56 measurements that pass the equilibration checks at the later cutoff. The retained sizes range from $L=64$ to 1024. The stiffness estimator in Eq.~\eqref{eq:rho} remains usable when the charge-$2e$ correlation ratio is unresolved. For each trial $T$ between 4.4325 and 4.4475, we interpolate $\rho_L$ linearly between measurements at most 0.02 apart.

At fixed $T$ we fit one parameter, $C$, by minimizing the inverse-variance-weighted sum of squared residuals $\chi^2(T,C)$ for $\rho_L/T$. We locate $T_4$ at the minimum of the resulting temperature profile. Figure~\ref{fig:stiffness}(b) plots $Q(T)=\min_C\chi^2(T,C)/(N_L-2)$, where $N_L$ is the number of sizes. This common normalization counts both $T_4$ and $C$ in the full fit. A fit at a prespecified temperature has $N_L-1$ residual degrees of freedom. The normalization leaves the location of the minimum unchanged. Each block-bootstrap sample repeats the minimization over $C$ and $T$ with the same interpolation and weights.

With successive time blocks merged in pairs, the WM minima occur at $T=4.4364$, 4.4365 and 4.4441 for $L_{\min}=64$, 128 and 256, respectively. Including the subleading logarithm gives 4.4381, 4.4371 and 4.4444. Across choices of cutoff, block size, data selection, size range and logarithmic form, the fitted centers span 4.4364--4.4444. At $L_{\min}=256$, the conditional 95\% bootstrap intervals span nearly the full temperature range searched. We therefore quote $T_4\simeq4.44$ conditional on half-vortex BKT scaling.

\section{Half-vortex pairs and domain-wall free energies}\label{sec:defects}

\subsection{Isolated half-vortex pairs}\label{sec:pairsector}

For an oriented triangular face, define its charge-$4e$ winding by $q_f=(2\pi)^{-1}\sum_{ij\in\partial f}\operatorname{Arg}e^{2i(\theta_j-\theta_i)}$. In $Z_{+-}(R,T;L)$ we fix the face with $q=+1$ and place the face with $q=-1$ at distance $R$ along a primitive lattice vector. Both faces have the same orientation. Every other face has $q=0$. This isolates one half-vortex pair and excludes screening by additional defects. The connecting domain wall fluctuates, and the periodic phases are integrated over all global windings. Fixing the endpoint vector removes the positional multiplicity of a radial shell.

In each measurement window the negative endpoint visits two prescribed faces, $s=a,b$. Denote the associated indicator function by $\mathbf1_s(\theta)$ and the fixed dimensionless sampling bias by $h_T(s)$. One replica has unnormalized weight
\begin{equation}
 \pi_T(\theta,s)=\mathbf1_s(\theta)e^{h_T(s)}
             \prod_{b'}w_{b'}(\delta_{b'},T),\qquad
 \Pi(\{\theta_j,s_j\})\propto\prod_{j=1}^{40}\pi_{T_j}(\theta_j,s_j).
\end{equation}
Here $w_b=\exp(-f_b/T)$ is the bond weight in Eq.~\eqref{eq:feff}. The 40 replica temperatures are geometrically spaced from the measurement temperature to 5. A local angle change can move the negative endpoint between the allowed faces. To move the endpoint collectively, we translate the phases by $\theta'_i=\theta_i+\theta_i^{s'}-\theta_i^s\pmod{2\pi}$, where $\theta^s$ and $\theta^{s'}$ are fixed reference configurations. The reverse translation has the same proposal probability and the Jacobian is one. A proposed state outside the allowed sectors has zero weight. For valid symmetric proposals, the Metropolis probability $\min[1,\pi_T(x')/\pi_T(x)]$ therefore satisfies detailed balance. Here $x=(\theta,s)$.

We supplement these moves with $\pi$-cluster flips, smooth phase and winding proposals, and occasional proposals drawing all phases uniformly. Adjacent replicas exchange complete configurations on alternating sets of temperature pairs. The exchange probability is
\[
 P_{\rm swap}=\min\!\left\{1,\,
 \frac{\pi_{T_j}(x_{j+1})\,\pi_{T_{j+1}}(x_j)}
      {\pi_{T_j}(x_j)\,\pi_{T_{j+1}}(x_{j+1})}\right\},
\]
with both the bond weights and the bias evaluated at the destination temperature.

For each window, two preliminary ladders determine the biases (16384 sweeps at $T=0.095$, 8192 otherwise). The biases are then fixed for four independent ladders started at the two endpoints, each with and without random local $\pi$ shifts. We sample the physical replica every eight sweeps. Its endpoint probabilities are proportional to $e^{h_T(s)}Z_s$, so the visit counts give $F_b-F_a=T[h_T(b)-h_T(a)-\ln(N_b/N_a)]$. In Fig.~4{}(a), the plotted $\Delta F_{1/2}$ is the sum of the direct increments $d_r=F(r+1)-F(r)$ starting at $r=2$. The direct windows are independent. Accordingly, the covariance of two cumulative values at $R$ and $R'$ is $\sum_{r=2}^{\min(R,R')-1}{\rm Var}(d_r)$.

We verify each direct increment by measuring the same free-energy difference in two independent steps through an intermediate face of opposite orientation. The two routes must agree within 3.5 combined standard errors. At both equilibration cutoffs, each independent run must visit each endpoint at least 50 times and complete at least four endpoint round trips and two temperature round trips. We require at least eight time blocks per run and an effective sample size of at least 100 for the slowest monitored observable. For occupation, energy and both windings, split $\widehat R$ must be at most 1.02. Independent starts must agree within 3.5 combined standard errors for the increment and four for energies and windings conditional on the endpoint. The shift between the two cutoffs must be within three times the larger standard error. For each increment we use the larger of the propagated four-start block-jackknife error and the standard error across starts. A curve enters Fig.~4{}(a) only when every window and both routes pass at both cutoffs.

\begin{table}[tbp]\centering
\caption{Direct increments $d_r$ from the later equilibration cutoff, in energy units; parentheses give one standard error. The $T=0.19$, $L=12$ windows did not pass the joint screen. The remaining $L=16$ temperatures were not sampled.}
\label{tab:pair}
\begin{tabular}{@{}ccccccc@{}}\toprule
 & $L=8$ & \multicolumn{2}{c}{$L=12$} & \multicolumn{3}{c}{$L=16$}\\
\cmidrule(lr){3-4}\cmidrule(l){5-7}
 $T$ & $d_2$ & $d_2$ & $d_3$ & $d_2$ & $d_3$ & $d_4$\\\midrule
 0.095 &17.248(7)&16.1724(37)&15.3539(47)&15.7854(60)&14.8840(50)&11.5626(38)\\
 0.19  &17.422(20)&---&---&---&---&---\\
 0.56  &17.872(38)&16.663(24)&14.775(19)&---&---&---\\
 1.20  &17.133(46)&15.538(34)&12.853(29)&---&---&---\\\bottomrule
\end{tabular}
\end{table}

At $T=0.095$, $L=8,12,16$, respectively, production reached 131072, 393216 and 655360 sweeps. The two cutoffs were $(65536,98304)$, $(131072,262144)$ and $(131072,393216)$. At $L=16$ the maximum closure discrepancy was 1.66 standard errors; the late measurement interval gave a minimum effective sample size of 427. At the three other temperatures the accepted series use 65536 sweeps with cutoffs 32768 and 49152. At $T=0.19,L=12$ all six windows were extended to 131072 sweeps and reexamined after cutoffs 65536 and 98304. The auxiliary step from the first to the intermediate face showed a 3.87-standard-error difference between endpoint starts at the earlier cutoff, falling to 0.62 at the later one. All closure discrepancies were below 1.58 standard errors. The size is excluded by the two-cutoff rule.

An open wall with asymptotic string tension $\tau_{\rm pair}$ would give $[F(2R)-F(R)]/R=\tau_{\rm pair}+A(T)\ln(2)/R+\cdots$. We keep the upper endpoint at $2R\le L/4$ for this comparison. Only the $T=0.095,L=16$ measurement supplies a qualifying doubling, $R=2\to4$, with slope $15.3347(39)$. A single finite-distance slope cannot separate a logarithmic attraction from a linear term.

\subsection{System-spanning domain walls}\label{sec:wallsector}

On the defect-free torus the winding of the charge-$4e$ field along primitive cycle $\gamma_\mu$ is $W_\mu=(2\pi)^{-1}\sum_{\gamma_\mu}\operatorname{Arg}e^{2i(\theta_j-\theta_i)}$. Project onto $\mathcal D_0$, the sector with $q_f=0$ everywhere and $W_x=W_y=0$. Periodic (P) boundaries have $\eta_b=1$ on every bond. For antiperiodic (AP) boundaries normal to cycle direction $\mu$, take $\eta_b=-1$ on all bonds crossing its seam, including diagonals. The two partition functions use the same projector:
\begin{equation}
 Z_{\eta}^{(0)}=\int_{\mathcal D_0}\!{\rm d}\mu(\theta)\,\sum_{\{n_b\}} e^{-H_\eta/T},\qquad
 H_\eta=\sum_b\big[-\eta_b(J_0+\lambda n_b)\cos\delta_b-J_2\cos2\delta_b+
                           (\varepsilon+\lambda)n_b\big].
\end{equation}
Vanishing plaquette and global windings allow a periodic real lift of $\phi$ with principal nearest-neighbor differences. Its square root has positive nearest-neighbor overlaps $c_b=\cos[(\phi_i-\phi_j)/2]$. A smooth twist of $\theta$ by $\pi$ around a cycle would wind $\phi$ by $2\pi$ and leave $\mathcal D_0$. Thus the seam probes the two charge-$2e$ branches at fixed global winding.

For fixed $\phi$ let $w(c)$ be the traced bond weight as a function of $\cos\delta$. Writing $e^{i\theta_i}=\sigma_i e^{i\phi_i/2}$ gives the dimensionless ferromagnetic coupling $k_b=\tfrac12\ln[w(c_b)/w(-c_b)]$. The common bond prefactor $\sqrt{w(c_b)w(-c_b)}$ is independent of $\eta_b$. For $\eta_b=\pm1$, the identity
\begin{equation}
 e^{k_b\eta_b\sigma_i\sigma_j}=e^{-k_b}
 \left[1+(e^{2k_b}-1)\mathbf1_{\sigma_i=\eta_b\sigma_j}\right]
\end{equation}
gives the Fortuin--Kasteleyn (FK) graph expansion. Given the periodic spins, an aligned edge is occupied with probability $1-e^{-2k_b}$, an antialigned edge with probability zero. Summing the spins gives two sign choices per connected component provided every occupied cycle has $\prod_b\eta_b=+1$. Periodic boundaries allow every graph. Antiperiodic boundaries exclude any graph with a cycle that crosses the seam an odd number of times.

Let $B_\mu$ indicate that a graph is compatible with the seam in direction $\mu$. Integrating the graph weights over $\phi$ gives
\begin{equation}
 p_\mu=\frac{Z_{{\rm AP},\mu}^{(0)}}{Z_{\rm P}^{(0)}}
 =\frac{\langle\mathbf1_{\mathcal D_0} B_\mu\rangle_{\rm P}}
        {\langle\mathbf1_{\mathcal D_0}\rangle_{\rm P}},\qquad
 \Delta F_{\rm wall}=-T\ln[(p_x+p_y)/2].
\end{equation}
We generate ordinary periodic phase histories with the updates of Sec.~\ref{sec:MC}. At each saved configuration we test the face charges and windings, drawing FK graphs only inside $\mathcal D_0$. Thus the projection enters the measurement. In the isolated-pair calculation of Sec.~\ref{sec:pairsector}, the prescribed charges constrain every update. On a $2\times2$ torus the FK estimator reproduces exact spin sums for both seam orientations and eight charge-$4e$ phase configurations.

\begin{table}[tbp]\centering
\caption{Total projected domain-wall free energy $\Delta F_{\rm wall}$ at the later cutoff, with one standard error in parentheses. A dagger marks insufficient sampling of a rare event at either cutoff. A dash indicates an unsampled size.}
\label{tab:wall}
\begin{tabular}{@{}lrrrrrr@{}}\toprule
 $T$ & $L=8$ & $16$ & $32$ & $64$ & $96$ & $128$\\\midrule
 0.095&0.2011(28)&0.3228(50)&0.594(12)&$\dagger$&---&---\\
 0.19&0.1163(20)&0.0561(12)&0.01112(37)&0.000293(37)&$\dagger$&---\\
 0.34&---&---&0.3138(16)&0.2590(15)&0.2110(13)&0.1674(11)\\
 0.37&---&---&0.4839(22)&0.5582(27)&0.6473(31)&0.7326(34)\\
 0.56&0.809(10)&1.007(14)&1.472(14)&2.445(23)&3.502(27)&4.619(24)\\
 1.20&0.3959(84)&0.0766(31)&0.00183(33)&$\dagger$&---&---\\\bottomrule
\end{tabular}
\end{table}

At $T=0.095,0.19,0.56,1.2$, the $L=8,16,32,64$ measurements use 4, 4, 8 and 16 independent chains, respectively. Each chain runs for 32768 sweeps, with cutoffs of 16384 and 24576 sweeps and two FK graphs every eight sweeps. At $T=0.34,0.37$ and for all sizes above 64, we use 16 chains and four FK graphs every eight sweeps. These chains run for 65536 sweeps with cutoffs of 32768 and 49152, except at $T=0.56,L=128$. There we extend the same 16 chains to 262144 sweeps and use only measurements after sweep 98304 or 131072 for the two estimates. All retained configurations were in $\mathcal D_0$.

We bootstrap time blocks within each chain (512 draws), keeping both seam directions and all graphs on a configuration together. We require at least eight blocks per chain and split $\widehat R\le1.03$. Each event type, seam-compatible and seam-incompatible, must occur in at least 24 blocks overall and two blocks in every chain, counting both directions together. The two seam estimates must agree within four standard errors, evaluated from the block variance of $B_x-B_y$. Runs with four FK graphs per configuration also require a within-chain ESS of at least 100 for every monitored observable. We apply these criteria at both cutoffs.

Table~\ref{tab:rare-wall} gives the counts separately for the two seams. If either event type is absent, we report no free-energy estimate. At $T=0.19,L=96$, incompatible graphs occur in only 12 blocks, so this point also fails the rare-event criterion. At $T=0.56,L=128$, the extended runs give $\Delta F_{\rm wall}=4.633(21)$ and $4.619(24)$ at the earlier and later cutoffs. At the later cutoff, every chain has at least 25 blocks containing a compatible graph, and the difference between the two seam free energies is $0.010(50)$.

\begin{table}[htbp]\centering\small
\caption{FK event counts after the later cutoff at selected temperatures and sizes. For each of the $N$ graphs, we test both seams. The counts $n_x,n_y$ refer to the less frequent event, specified in column three. The last column gives the smallest within-chain ESS among the monitored quantities (ratio estimator, sector occupancy, order-parameter moments, and difference between seams).}
\label{tab:rare-wall}
\begin{tabular}{@{}rrlrrrr@{}}\toprule
 $T$ & $L$ & Event & $N$ & $n_x$ & $n_y$ & Min.\ ESS \\\midrule
 0.095 & 64  & Compatible   & 32768  & 0   & 0   & 205\\
 1.20  & 64  & Incompatible & 32768  & 0   & 0   & 414\\
 0.19  & 96  & Incompatible & 131072 & 8   & 4   & 543\\
 0.34  & 32  & Compatible   & 131072 & 51985 & 52191 & 845\\
 0.34  & 96  & Incompatible & 131072 & 60539 & 60662 & 653\\
 0.37  & 32  & Compatible   & 131072 & 35340 & 35543 & 821\\
 0.37  & 96  & Compatible   & 131072 & 22969 & 22612 & 679\\
 0.56  & 96  & Compatible   & 131072 & 238 & 266 & 634\\
 0.56  & 128 & Compatible   & 1048576 & 272 & 277 & 6162\\\bottomrule
\end{tabular}
\end{table}

On opposite sides of $T_2$, the total free energy falls at $T=0.34$ and rises at $T=0.37$ at each step from $L=32$ to 64, 96 and 128 (Table~\ref{tab:wall}). Both cutoffs give these trends. At $T=0.56$, the finite-size slope $[\Delta F_{\rm wall}(128)-\Delta F_{\rm wall}(96)]/32$ is $0.0349(11)$. The corresponding values of $\Delta F_{\rm wall}/L$ are $0.03648(28)$ at $L=96$ and $0.03608(19)$ at $L=128$. Within the precision of these two sizes the total cost is compatible with linear growth. A positive projected interfacial tension $\tau_{\rm wall}^{(0)}$ would require $\Delta F_{\rm wall}/L$ to approach a positive limit. The string tension $\tau_{\rm pair}$ is defined separately by increasing the endpoint separation in the isolated-pair ensemble.

\section{Three-branch Potts model}\label{sec:potts}
Replacing the second harmonic in main-text Eq.~\textup{(2)} by
\(-J_3\cos3\delta_b\) changes the number of allowed branches in the constrained
limit. At \(J_3\to\infty\) and fixed \(T\), \(e^{3i\theta_i}\) is uniform and
\(\theta_i=\theta_0+2\pi s_i/3\) with \(s_i=0,1,2\).
For equal and unequal neighboring branches, \(\cos\delta_b\) is \(1\) and
\(-1/2\), respectively. Up to the common third-harmonic factor the bond weights are
\begin{equation}
 W_{=}=e^{J_0/T}(1+e^{-\varepsilon/T}),\qquad
 W_{\ne}=e^{-J_0/(2T)}(1+e^{-(\varepsilon+3\lambda/2)/T}).
\end{equation}
The resulting three-state Potts weight is
\(\exp[\mathcal K_3(T)\sum_{\langle ij\rangle}\delta_{s_i,s_j}]\), with
\begin{equation}
 \mathcal K_3=\ln(W_{=}/W_{\ne})
    =\frac{3J_0}{2T}+\ln\frac{1+e^{-\varepsilon/T}}
                                 {1+e^{-(\varepsilon+3\lambda/2)/T}}.
 \label{eq:pottsmap}
\end{equation}
Here the dimensionless Potts coupling multiplies a Kronecker delta. On the triangular
lattice the ferromagnetic critical point satisfies \(v_c^3+3v_c^2=3\),
where \(v_c=e^{\mathcal K_{3,c}}-1\)~\cite{WuGuo2012}. Hence
\(\mathcal K_{3,c}=\ln[2\cos(\pi/9)]\simeq0.631\).
Writing \(p_{=}\) and \(p_{\ne}\) for the excitation probabilities of the two
bond types gives \(\mathcal K_3+T\mathcal K_3'=s(p_{=})-s(p_{\ne})<\ln2\).
Thus the triangular threshold lies below the available entropy bound.

To establish an open reentrant region, fix \(\lambda>0\), set
\(\varepsilon=d\lambda\), \(J_0=d^3\lambda\), and take \(d>0\) small.
For each such choice the microscopic couplings remain fixed on heating.
At \(T_a=d^2\lambda\) both excited-state Boltzmann factors vanish exponentially as
\(d\to0\), and \(3J_0/(2T_a)\) tends to zero. At \(T_b=\sqrt d\,\lambda\),
the excited-state Boltzmann factors \(e^{-\varepsilon/T_b}\) and
\(e^{-(\varepsilon+3\lambda/2)/T_b}\) tend to one and zero for equal and unequal
branches, respectively. The corresponding excitation probabilities tend to \(1/2\) and zero. Thus
\(\mathcal K_3(T_a)\to0\) and \(\mathcal K_3(T_b)\to\ln2\).
For sufficiently small fixed \(d\), \(T_a<T_b\) and the corresponding
couplings lie on opposite sides of \(\mathcal K_{3,c}\).
Since \(\mathcal K_3(0^+)=\infty\) and \(\mathcal K_3(\infty)=0\),
continuity gives at least three crossings. One can choose four finite
temperatures, one in each successive interval, at which \(\mathcal K_3\)
lies strictly above, below, above, and below threshold. These inequalities
persist under small changes in the couplings. Reentrant Potts branch order
therefore occurs in an open parameter region of the constrained model.
Whether charge-\(6e\) superconductivity reenters at finite \(J_3\) remains open.

\paragraph{Data and code availability.}
The simulation data and analysis code are available from the authors upon reasonable request.
\begingroup\small\setlength{\bibsep}{2pt}
\makeatletter
\let\SMoriginalbibitem\@lbibitem
\def\@lbibitem[#1]#2{%
  \ifcsname SMbibnumber@#2\endcsname
    \setcounter{NAT@ctr}{\numexpr\csname SMbibnumber@#2\endcsname-1\relax}%
  \fi
  \SMoriginalbibitem[#1]{#2}}
\newcommand{\SMbibnumber}[2]{\expandafter\def\csname SMbibnumber@#1\endcsname{#2}}
\makeatother
\SMbibnumber{Shi2011}{38}
\SMbibnumber{Serna2017}{40}
\SMbibnumber{Houtappel1950}{49}
\SMbibnumber{Liaw1999}{50}
\SMbibnumber{Wu2006}{51}
\SMbibnumber{Nelson1977}{52}
\SMbibnumber{Weber1988}{53}
\SMbibnumber{Hsieh2013}{54}
\SMbibnumber{Swendsen1987}{55}
\SMbibnumber{Onsager1944}{56}
\SMbibnumber{WuGuo2012}{57}
\bibliographystyle{apsrev4-2}\bibliography{references}\endgroup